\documentstyle[12pt,epsfig,epsf,rotating,axodraw]{article}

\newcommand{\lsim}{\raisebox{-0.13cm}{~\shortstack{$<$ \\[-0.07cm] $\sim$}}~}
\newcommand{\gsim}{\raisebox{-0.13cm}{~\shortstack{$>$ \\[-0.07cm] $\sim$}}~}

\newcommand{\ra}{\rightarrow}
\newcommand{\ee}{e^+e^-}
\newcommand{\s}{\\ \vspace*{-3.5mm} }
\newcommand{\nn}{\noindent}
\newcommand{\non}{\nonumber}
\newcommand{\beq}{\begin{eqnarray}}
\newcommand{\eeq}{\end{eqnarray}}
\newcommand{\tb}{\tan\beta}
\newcommand{\dx}{{\rm d}}
\newcommand{\ct}[1]{c_{\theta_#1}}
\newcommand{\st}[1]{s_{\theta_#1}}
\begin{document}
\baselineskip=18pt

\def\thefootnote{\fnsymbol{footnote}}

\begin{flushright}
PM/01--20\\
March 2001\\
\end{flushright}

\vspace{1cm}

\begin{center}

{\large\sc {\bf Chargino and Neutralino Decays Revisited}}

\vspace{1cm}

{\sc A. Djouadi$^1$, Y. Mambrini$^{1,2}$} and {\sc M. M\"uhlleitner$^1$}

\vspace{0.5cm}

$^1$ Laboratoire de Physique Math\'ematique et Th\'eorique, UMR5825--CNRS,\\
Universit\'e de Montpellier II, F--34095 Montpellier Cedex 5, France. 

\vspace{0.3cm}

$^2$ CEA/DIF/DPTA/SPN, B.P. 12, F--91680 Bruy\`eres le Ch\^atel, France.  

\end{center}

\vspace{2cm}

\begin{abstract}
\nn We perform a comprehensive analysis of the decays of charginos and
neutralinos in the Minimal Supersymmetric Standard Model where the neutralino
$\chi_1^0$ is assumed to be the lightest supersymmetric particle. We focus, in
particular, on the three--body decays of the next--to--lightest neutralino and
the lightest chargino into the lightest neutralino and fermion--antifermion
pairs and include vector boson, Higgs boson and sfermion exchange diagrams,
where in the latter contribution the full mixing in the third generation is
included. The radiative corrections to the heavy fermion and SUSY particle
masses will be also taken into account. We present complete analytical formulae
for the Dalitz densities and the integrated partial decay widths in the massless
fermion case, as well as the expressions of the differential decay widths
including the masses of the final fermions and the polarization of the decaying
charginos and neutralinos.  We then discuss these decay modes, in particular 
in scenarios where the parameter $\tan \beta$ is large and in models without 
universal gaugino masses at the Grand Unification scale, where some new decay 
channels, such as decays into gluinos and $q\bar{q}$ pairs,  open up. 
\end{abstract}

\def\thefootnote{\arabic{footnote}}
\setcounter{footnote}{0}
\baselineskip=18pt
\newpage

\section*{1. Introduction}

In the Minimal Supersymmetric Standard Model (MSSM) \cite{R1,Haber-Kane}, the
lightest neutralinos $\chi_{1}^0, \chi_{2}^0$ and chargino $\chi_1^\pm$, which
are mixtures of the higgsinos and gauginos that are the spin $\frac{1}{2}$
partners of the Higgs and gauge bosons, are expected to be the lightest
supersymmetric particles. In particular, the neutralino $\chi_1^0$ is the
lightest SUSY particle (LSP), which because of R--parity conservation
\cite{R2}, is stable and invisible.  In models where the gaugino masses are
unified at the Grand Unification scale \cite{mSUGRA}, the masses of these
particles are such that: $m_{\chi_2^0} \sim m_{\chi_1^\pm} \sim 2 m_{\chi_1^0}$
in the case where they are gaugino--like or $m_{\chi_2^0} \sim m_{\chi_1^\pm}
\sim m_{\chi_1^0}$ in the case where they are higgsino--like.  Thus, the states
$\chi_2^0$ and $\chi_1^+$ are not much heavier than the LSP and might be the
first SUSY particles to be discovered. The search for these sparticles is a
major goal of present and future colliders, and the detailed study of their
production and decay properties is mandatory in order to reconstruct the SUSY
Lagrangian at the low energy scale and to derive the structure of the theory at
the high scale. \s

The decays of charginos and neutralinos have been widely discussed in the
literature \cite{R3}. If the mass splitting between the LSP and the
next--to--lightest neutralino $\chi_2^0$ or the lightest chargino $\chi_1^\pm$
is larger than $M_Z$ or $M_W$, the particles will decay into massive gauge
bosons and the neutralino $\chi_1^0$. If not, the decays will occur through
virtual gauge boson and scalar fermion exchanges, leading in the final state to
the LSP neutralino and a fermion--antifermion pair. Recently, it has been
realized [6--9] that for large values of the parameter $\tb$, the
ratio of the vacuum expectation values of the two doublet Higgs fields which
are needed to break the electroweak symmetry in the MSSM, the Yukawa couplings
of third generation down--type fermions [$b$ quarks and $\tau$ leptons], which
are strongly enhanced, lead to dramatic consequences for the decays of these
particles\footnote{Note that the scenario with $\tb \sim m_t/m_b$ is favored in
models with Yukawa coupling unification at the GUT scale \cite{R6}. In
addition, large $\tb$ values, $\tb \gsim 3$--8 depending on the details of the
radiative corrections, are needed to maximize the lightest $h$ boson mass in
the MSSM, to cope with the LEP2 experimental bound $M_h \gsim 113.5$ GeV
\cite{R7} in the decoupling regime where the $h$ boson is Standard Model
like.}.  Indeed, the virtual exchanges of, on the one side, Higgs particles
[because the Higgs boson couplings to $b$ quarks and $\tau$ leptons are
proportional to $\tb$] and, on the other side, of third generation down--type
sfermions [which tend to be lighter than the other sfermions  in this case]
become very important. \s

Furthermore, some interest has been recently devoted to models where the
gaugino masses are not unified at the GUT scale, as it might be the case in a
large class of four--dimensional string models \cite{R8a} or in the so--called
anomaly--mediated SUSY breaking models \cite{R8b}. As an example, two
particular cases have been discussed in Ref.~\cite{R8}, where SUSY--breaking
occurs via an F--term that is not an SU(5) singlet and in an orbifold string
model.  In these models the gaugino masses at the electroweak scale can be very
different from the pattern mentioned above. In particular, the $\chi_2^0$ and
$\chi_1^\pm$ masses can be closer to the LSP mass in some of these models,
favoring the occurrence of three--body decays of the light chargino and
neutralino states [including some some new channels such as $\chi_2^0 \to
q\bar{q}+$gluino final states], while possibly disfavoring final states with
heavy fermions [such as $b\bar{b}$ final states] and therefore affecting
dramatically the decay branching ratios. \s

In this paper, we perform a detailed investigation of the three--body decay
modes of charginos and neutralinos in the MSSM, focusing on the scenarios with
large values of $\tb$ and with non--unified gaugino masses at the GUT scale. We
will provide complete analytical formulae for the Dalitz densities of the
decays [in terms of the energies of the two final state fermions] and for the
fully integrated partial decay widths. Furthermore, we will take into account
the polarization of the decaying particle, which is needed in order to obtain
the full correlations between the initial state in the production of these
particles and the final states in their decays. We will also include the
dependence on the masses of the final state fermions to have a more accurate 
prediction for final states involving $b$--quarks and $\tau$ leptons 
[especially in scenarios where the mass difference between the decaying 
particles and the LSP is not very large] and to treat properly the case of  
heavy top quark final states. An important ingredient of the analysis will be 
the inclusion of the effects of the radiative corrections to the heavy fermion
and chargino/neutralino masses, which will be shown to have a large impact. \s
  
This work extends on the recent analyses made in Refs.~[6--8] for
chargino and neutralino decays, and completes our analyses of the higher order
decays of SUSY particles [sfermions, in particular stops and sbottoms, and 
gluinos] in the MSSM \cite{N1,N2}. \s

The paper is organized as follows. In the next section, we will summarize the
main features of the chargino, neutralino, sfermion and Higgs sectors of the
MSSM which will be needed in our analysis. In section 3, we will display the
analytical expressions of the (unpolarized) Dalitz densities and the integrated
partial three--body decay widths for massless final state fermions. Section 4
will be devoted to our numerical analysis and a short conclusion will be given
in section 5. In the Appendix, we present the complete formulae for the partial
decay widths, including the finite mass of the fermion final states and the
polarization of the decaying charginos and neutralinos.  

\section*{2. SUSY particles masses and couplings}
\setcounter{equation}{0}
\renewcommand{\theequation}{2.\arabic{equation}}

To fix our notation, we will summarize in this section the main features of the
chargino, neutralino, sfermion and Higgs sectors of the MSSM.  We will then
give, for completeness, all the couplings of these SUSY particles [i.e. 
couplings of the neutralinos and charginos to gauge and Higgs bosons and their
couplings to fermion--sfermion pairs] as well as the couplings of MSSM Higgs
and gauge bosons to fermions, which will be needed later when evaluating the
two--body and three--body partial decay widths.  

\subsection*{2.1 Masses and mixing}

\subsubsection*{2.1.1 The chargino and neutralino systems} 

The general chargino mass matrix, in terms of the wino mass parameter
$M_2$, the higgsino mass parameter $\mu$ and $\tb$, is given by \cite{R9}
\begin{eqnarray}
{\cal M}_C = \left[ \begin{array}{cc} M_2 & \sqrt{2}M_W s_\beta
\\ \sqrt{2}M_W c_\beta & \mu \end{array} \right]
\end{eqnarray}
where we use $s_\beta \equiv \sin \beta \, , \, c_\beta \equiv \cos 
\beta$ etc.. It is diagonalized by two real matrices $U$ and $V$, 
\begin{eqnarray}
U^* {\cal M}_C V^{-1} \ \ \ra \ \ U={\cal O}_- \ {\rm and} \ \ V = 
\left\{
\begin{array}{cc} {\cal O}_+ \ \ \ & {\rm if \ det}{\cal M}_C >0  \\
            \sigma_3  {\cal O}_+ \ \ \ & {\rm if \ det}{\cal M}_C <0  
\end{array}
\right. 
\end{eqnarray}
where $\sigma_3$ is the Pauli matrix to make the chargino masses positive and 
${\cal O}_\pm$ are rotation matrices, with angles given by:
\begin{eqnarray}
\tan 2 \theta_- =  \frac{ 2\sqrt{2}M_W(M_2 c_\beta
+\mu s_\beta)}{ M_2^2-\mu^2-2M_W^2 c_\beta} \ \ , \ \ 
\tan 2 \theta_+ = \frac{ 2\sqrt{2}M_W(M_2 s_\beta
+\mu c_\beta)}{M_2^2-\mu^2 +2M_W^2 c_\beta} 
\end{eqnarray}
This leads to the two chargino masses:
\begin{eqnarray}
m^2_{\chi_{1,2}^\pm} = \frac{1}{2} \left\{ M_2^2+\mu^2+2M_W^2
\mp \left[ (M_2^2-\mu^2)^2+4 M_W^2( M_W^2 c^2_{2\beta} + M^2_2+\mu^2
+2M_2\mu s_{2\beta}) \right]^{\frac{1}{2}} \right\} 
\end{eqnarray}
In the limit $|\mu| \gg M_2, M_W$, the masses of the two charginos reduce to
\begin{eqnarray}
m_{\chi_{1}^\pm}  \simeq   M_2 - \frac{M_W^2}{\mu^2} 
\left( M_2 +\mu s_{2\beta} \right) \ \ , \ \ 
m_{\chi_{2}^\pm}  \simeq  |\mu| + 
\frac{M_W^2}{\mu^2} \epsilon_\mu \left( M_2 s_{2 \beta} +\mu \right) 
\end{eqnarray}
where $\epsilon_\mu$ is for the sign of $\mu$. For $|\mu| \ra \infty$,
the lightest chargino corresponds to a pure wino state with mass 
$m_{\chi_{1}^\pm} \simeq M_2$, while the heavier chargino corresponds to a 
pure higgsino state with a mass $m_{\chi_{2}^\pm} = |\mu|$. \smallskip

In the case of the neutralinos, the four--dimensional neutralino mass matrix 
depends on the same two mass parameters $\mu$ and $M_2$, if the GUT 
relation $M_1=\frac{5}{3} \tan^2 \theta_W$, $M_2 \simeq \frac{1}{2} M_2$ 
\cite{R9} is used. In the $(-i\tilde{B}, -i\tilde{W}_3, \tilde{H}^0_1,$ 
$\tilde{H}^0_2)$ basis, it has the form [$c_W^2=1-s_W^2=M_W^2/M_Z^2$]  
\begin{eqnarray}
{\cal M}_N = \left[ \begin{array}{cccc}
M_1 & 0 & -M_Z s_W c_\beta & M_Z  s_W s_\beta \\
0   & M_2 & M_Z c_W c_\beta & -M_Z  c_W s_\beta \\
-M_Z s_W c_\beta & M_Z  c_W c_\beta & 0 & -\mu \\
M_Z s_W s_\beta & -M_Z  c_W s_\beta & -\mu & 0
\end{array} \right]
\end{eqnarray}
It can be diagonalized analytically \cite{R11} by a single real matrix $Z$. 
The expressions of the masses $m_{\chi_i^0}$ are rather involved. In the 
limit of large $|\mu|$ values, they however simplify to \cite{R10} 
\begin{eqnarray}
m_{\chi_{1}^0} &\simeq& M_1 - \frac{M_Z^2}{\mu^2} \left( M_1 +\mu s_{2\beta}
\right) s_W^2 \non \\
m_{\chi_{2}^0} &\simeq& M_2 - \frac{M_Z^2}{\mu^2} \left( M_2 +\mu s_{2 \beta}
\right) c_W^2 \non \\
m_{\chi_{3}^0} &\simeq& |\mu| + \frac{1}{2}\frac{M_Z^2}{\mu^2} \epsilon_\mu 
(1- s_{2\beta}) \left( \mu + M_2 s_W^2+M_1 c_W^2 \right) \non \\
m_{\chi_{4}^0} &\simeq& |\mu| + \frac{1}{2}\frac{M_Z^2}{\mu^2} \epsilon_\mu 
(1+s_{2\beta}) \left( \mu - M_2 s_W^2 - M_1 c_W^2 \right) 
\end{eqnarray}
Again, for $|\mu| \ra \infty$, two neutralinos are pure gaugino states 
with masses $m_{\chi_{1}^0} \simeq M_1$, $m_{\chi_{2}^0} =M_2$, while
the two others are pure higgsino states, with masses $m_{\chi_{3}^0} \simeq 
m_{\chi_{4}^0} \simeq |\mu|$. The matrix elements of the diagonalizing matrix, 
$Z_{ij}$ with $i,j=1,..4$, are given by
\begin{eqnarray}
Z_{i1} &=& \left[ 1+ \left( \frac{Z_{i2}}{Z_{i1}} \right)^2+\left( \frac{Z_{i3}}
{Z_{i1}} \right)^2+\left( \frac{Z_{i4}}{Z_{i1}} \right)^2 \right]^{-1/2} \\
\frac{Z_{i2}}{Z_{i1}} &=&  -\frac{1}{\tan \theta_W} \frac{ M_1-
\epsilon_i  m_{\chi_{i}^0} } {M_2 -\epsilon_i m_{\chi_{i}^0} } \non \\
\frac{Z_{i3}}{Z_{i1}} &=&  \frac{
\mu (M_1-\epsilon_i  m_{\chi_{i}^0} )(M_2 -\epsilon_i m_{\chi_{i}^0} )
-M_Z^2 s_\beta c_\beta [(M_1-M_2)c_W^2+M_2 -\epsilon_i m_{\chi_{i}^0}] } 
{M_Z (M_2 -\epsilon_i m_{\chi_{i}^0} )s_W [\mu c_\beta 
+\epsilon_i m_{\chi_{i}^0} s_\beta) } \non \\
\frac{Z_{i4}}{Z_{i1}} &=&  \frac{
-\epsilon_i m_{\chi_{i}^0} (M_1-\epsilon_i m_{\chi_{i}^0} )(M_2 -\epsilon_i 
m_{\chi_{i}^0} ) -M_Z^2 c^2\beta [(M_1-M_2)c_W^2+M_2 
-\epsilon_i  m_{\chi_{i}^0}] } 
{M_Z (M_2 -\epsilon_i m_{\chi_{i}^0} )s_W [\mu c_\beta
+\epsilon_i m_{\chi_{i}^0} s_\beta) } \non 
\end{eqnarray}
where $\epsilon_i$ is the sign of the $i$th eigenvalue of the neutralino 
mass matrix, which in the large $|\mu|$ limit are: $\epsilon_1
=\epsilon_2=1$ and $\epsilon_4=-\epsilon_3=\epsilon_\mu$. Note that we will 
often use the rotated $Z_{ij}$ matrix elements: 
\begin{eqnarray}
Z'_{i1}=Z_{i1}c_W +Z_{i2} s_W \ , \ 
Z'_{i2}=-Z_{i1} s_W + Z_{i2} c_W \ , \ 
Z'_{i3}=Z_{i3} \  , \ 
Z'_{i4}=Z_{i4} 
\end{eqnarray} 

We will not only discuss the chargino and neutralino spectrum in mSUGRA--type 
models, where the gaugino masses are unified at the GUT scale $M_{\rm GUT}$, 
but also when the boundary conditions at this high scale are different.
For illustration, we focus on two scenarios discussed in Ref.~\cite{R8}: $i)$
Models in which SUSY breaking occurs via an F--term that is not SU(5) singlet
but belongs to a representation which appears in the symmetric product of two
adjoints: ({\bf 24}$\otimes${\bf 24})$_{\rm sym}$={\bf 1}$\oplus${\bf
24}$\oplus${\bf 75}$\oplus${\bf 200} [where only model {\bf 1} leads to the
universal gaugino masses discussed previously].  $ii)$ The {\bf OII} model
which is superstring motivated and where the SUSY breaking is
moduli--dominated. \smallskip 

The relation between the gaugino masses at the scale $M_{\rm GUT}$, 
$m_{1,2,3}$, and at the weak scale ${\cal O}(M_Z)$, $M_{1,2,3}$, are 
approximately given by the relation \cite{Komine}: 
\beq
M_1 \simeq 0.42 m_1 \ , \ M_2 \simeq 0.83 m_2 \ , \  M_3 \simeq 2.6 m_3 
\eeq
leading to the well known hierarchy $M_1:M_2:M_3=1:2:6$ for a universal gaugino
mass at the GUT scale, $m_1=m_2=m_3=m_{1/2}$, as in mSUGRA type models.  The
relative gaugino masses at $M_{\rm GUT}$ and at the low--energy scale $M_Z$ are
given in Table 1; Ref.~\cite{R8}. The pattern for the neutralino and chargino
masses can be quite different from the universal case {\bf 1}. In particular,
for large values of the parameter $\mu$, the LSP is wino--like in the scenario
{\bf 200} where $M_2 <M_1$, implying that $\chi_1^0$ and $\chi_1^+$ are
degenerate in mass. In the scenario {\bf 75}, the gauginos $\chi_1^0, \chi_2^0$
and $\chi_1^+$ have masses which are very close since $|M_1| \sim |M_2|$, while
in scenario {\bf 24}, the mass splitting between the LSP and the states
$\chi_2^0,\chi_1^+$ can be very large.  In the {\bf OII} model and if no large
loop corrections are present to increase the gluino mass compared to the value
of $M_3 <M_1, M_2$ [to avoid the scenario with a gluino LSP], $\chi_1^0,
\chi_2^0$ and $\chi_1^+$ have to be higgsino like and can be thus degenerate in
mass.

\begin{table}[htbp!]
\begin{center}
\renewcommand{\arraystretch}{1.5}
\begin{tabular}{|c||c|c|c|} \hline
$\ \ \ F_\Phi \ \ \ $ & $\ \ M_3 \ \ $ & $\ \ M_2 \ \ $ & $\ \ M_1\ \ $
\\ \hline
{\bf 1} & $1 (\sim 6)$ & $1(\sim 2)$ & $1 (\sim 1)$ \\ \hline
{\bf 24} & $2 (\sim 12)$ & $-3 (\sim -6)$ & $-1 (\sim -1)$ \\ \hline
{\bf 75} & $1 (\sim 6)$ & $3 (\sim 6)$ & $-5 (\sim -5 )$ \\ \hline
{\bf 200} & $1 (\sim 6)$ & $2 (\sim 4)$ & $10 (\sim 10)$ \\ \hline \hline
{\bf OII} & $1 (\sim 6)$ & $5 (\sim 10)$ & $53/5 (\sim 53/5)$ 
\\ \hline
\end{tabular}
\end{center}
\caption{\small \it Relative gaugino masses at $M_{\rm GUT} (M_Z)$ 
in the $F_\Phi$ representations and the OII model.}
\end{table}

Since $\chi_2^0$ and $\chi_1^+$ can be degenerate in mass with the LSP
in some of these scenarios, it is important to include the radiative
corrections to the masses. These corrections are quite involved \cite{radchi}.
Here we will work in two different approximations, which are valid in
the (almost) pure gaugino and pure higgsino regions \cite{Radcor,Manuel} and
which reproduce the complete result to better than a few percent. \s 

For gaugino like neutralinos and charginos, $|\mu| \gg M_{1}, M_2,M_Z$, we will
correct only the parameters, $M_1, M_2$ in the chargino and neutralino mass
matrices [which means that terms of ${\cal O}(\alpha/4\pi \times M_Z^2/ \mu^2)$
are neglected]; we assume that all fermions are massless and all squarks and
sleptons are degenerate, with masses $m_{\tilde{q}}$ and $m_{\tilde{l}}$,
respectively; furthermore we work in the tree--level decoupling limit for the
Higgs sector, where $M_h \sim M_Z$ and $M_{H} \sim M_{H^+} \sim M_A$  [see
section 2.1.3]. For the gluino mass, $m_{\tilde{g}}=M_3+\Delta M_3/M_3$, needed
in order to compare to the LSP mass, we will include only the dominant QCD
corrections.  \s

In this limit, one then obtains for $\Delta M_{1,2,3}/ M_{1,2,3}$ 
\cite{Radcor}:  
\beq
\frac{\Delta M_1}{M_1} &=& - \frac{\alpha}{4\pi c_W^2} \left\{ 
11 B_1(M_1^2, 0, m_{\tilde{q}})+ 9 B_1(M_1^2, 0, m_{\tilde{l}}) - \frac{\mu}
{M_1} s_{2\beta}  \right. \\
&& \left. \times \Bigg[ B_0(M_1^2, \mu, M_A) - B_0 (M_1^2, \mu, M_Z) \Bigg] +
B_1(M_1^2, \mu, M_A) + B_1 (M_1^2, \mu, M_Z) \right\} \non \\
\frac{\Delta M_2}{M_2} &=& - \frac{\alpha}{4\pi s_W^2} \left\{ 
9B_1(M_2^2, 0, m_{\tilde{q}})+ 3 B_1(M_1^2, 0, m_{\tilde{l}}) - \frac{\mu}
{M_2} s_{2\beta}  \right. \\
&& \times \Bigg[ B_0(M_2^2, \mu, M_A) - B_0 (M_2^2, \mu, M_Z) \Bigg] +
B_1(M_2^2, \mu, M_A) + B_1 (M_2^2, \mu, M_Z)  \non \\
&&  \left. - 8 B_0(M_2^2, M_2, M_W) + 4 B_1 (M_2^2, M_2, M_W) \right\} \non \\
\frac{\Delta M_3}{M_3} &=& \frac{3\alpha_s}{2\pi} \left\{ 
2B_0(M_3^2, M_3, 0)- B_1(M_3^2, M_3, 0) -2 B_1(M_3^2, 0, m_{\tilde{q}}) 
\right\}
\eeq
with the finite parts of the Passarino--Veltman two--point functions $B_1$ and 
$B_0$ given by \cite{PV}:
\beq
B_0(q^2, m_1,m_2) &=& -{\rm Log}\left(\frac{q^2}{Q^2} \right)-2  \non \\
&& -{\rm Log}(1-x_+)-x_+{\rm Log}(1-x_+^{-1}) 
-{\rm Log}(1-x_-)-x_-{\rm Log}(1-x_-^{-1}) \non \\
B_1(q^2, m_1,m_2) &=& \frac{1}{2q^2} \bigg[ m_2^2 \left(1- {\rm \log} 
\frac{m_2^2}{Q^2} \right) - m_1^2 \left(1- {\rm Log} \frac{m_1^2}{Q^2} 
\right)  \non \\ &&  + (q^2-m_2^2+m_1^2) B_0(q^2, m_1,m_2) \bigg]  
\eeq
with $Q^2$ denoting the renormalization scale and  
\beq
x_{\pm} = \frac{1}{2q^2} \left( q^2-m_2^2+m_1^2 \pm \sqrt{(q^2-m_2^2+m_1^2)^2 
-4q^2(m_1^2- i \epsilon) } \, \right)
\eeq 

For higgsino--like $\chi_1^0, \chi_2^0$ and $\chi_1^+$ particles, $|\mu| \ll 
M_{1,2}$, we will follow the approach of Ref.~\cite{Manuel} and only correct 
the higgsino entries in the neutralino mass matrix and include the dominant 
Yukawa corrections to the light chargino and neutralino masses, due to stop/top
and sbottom/bottom loops\footnote{We will further approximate the $\delta_C$ 
correction by $\delta_{34}$ in the paper \cite{Manuel}, which would be the case 
for almost degenerate squarks; the difference is negligible in general.}. 
The masses in the higgsino limit \cite{Manuel} are then given by [we keep the 
sign of the eigenvalues]: 
\beq
m_{\chi_1^\pm} &\simeq & |\mu + \delta_C| \left[ 1-\frac{M_W^2 s_{2\beta}}{ 
M_2 (\mu+\delta_C)} \right] \non \\
m_{\chi_{1,2}^0} &\simeq & \mp (\mu+ \delta_{C})- \frac{M_Z^2}{2} (1 \mp 
s_{2\beta} ) \left( \frac{s_W^2}{M_1^2}+\frac{c_W^2}{M_2^2} \right) +
\delta_N
\eeq
with
\beq
\delta_C &=& \frac{- 3 \alpha \mu }{8 \pi } \bigg[ \lambda_t^2 \left(
B_1(\mu^2, m_t, m_{\tilde{t}_1}) +B_1(\mu^2, m_t, m_{\tilde{t}_2}) \right)
\non \\
&& \hspace*{.8cm} + \lambda_b^2 \left(
B_1(\mu^2, m_b, m_{\tilde{b}_1}) +B_1(\mu^2, m_b, m_{\tilde{b}_2}) \right)
\bigg] \non \\
\delta_N &=& \frac{-3\alpha}{8 \pi } \bigg[ \lambda_t^2 \, m_t s_{2\theta_t} 
\,  \left( B_0(\mu^2, m_t, m_{\tilde{t}_1}) - B_0(\mu^2, m_t, m_{\tilde{t}_2}) 
\right) \non \\
&& \hspace*{.8cm} + \lambda_b^2 m_b s_{2 \theta_b} \,  \left(
B_0(\mu^2, m_b, m_{\tilde{b}_1}) -B_0(\mu^2, m_b, m_{\tilde{b}_2}) \right)
\bigg] 
\eeq
where $\theta_{t,b}$ are the mixing angles in the stop and sbottom sectors
[to be discussed in the next subsection] and $\lambda_{t,b}$ are the reduced
Yukawa couplings of the $t,b$ quarks, which in terms of the running masses 
[to be also discussed in the next subsection] are given by: 
\beq
\lambda_b = \frac{m_b}{\sqrt{2} M_W s_W c_\beta} \ \ , \ \ 
\lambda_t = \frac{m_t}{\sqrt{2} M_W s_W s_\beta} 
\eeq
The $\chi_1^\pm$ and $\chi_2^0$ masses as well as the mass differences
$m_{\chi_1^\pm}  - m_{\chi_1^0}$ and $m_{\chi_2^0}  - m_{\chi_1^0}$ are shown
in Fig.~1 as a function of $\mu$ for $\tan \beta=50$, in the five models
discussed above. The wino mass parameter is fixed to $M_2=150$ GeV and the
parameter $M_1$ is obtained from $M_2$ as in Table 1. We see that the mass
difference between the lightest chargino and the LSP can be very
small\footnote{The search for charginos and neutralinos, which are almost
degenerate in mass with the LSP, can be done in $\ee$ collisions, either via a
search of almost stable particles or by a search of multi--pion final states
with a large amount of missing energy; see for instance Ref.~\cite{searches}. 
At hadron colliders, the direct search of such states will be very difficult, 
if possible at all.} in models {\bf OII} and {\bf 200}, even after the inclusion
of the radiative corrections. In model ${\bf 75}$, the next--to--lightest
neutralino and the lightest chargino can be degenerate in mass with the LSP for
small values of $\mu$, and the mass difference hardly exceeds 20 GeV [for the
chosen value of $M_2$] even for large $\mu$ values. Note that for values
$\mu \gsim M_{3}$, the gluino is lighter than the lightest neutralino 
$\chi_1^0$ in model {\bf OII}.  

\begin{figure}[htbp]
\vspace*{-4.5cm}
\hspace*{-2.3cm}
\mbox{\psfig{figure=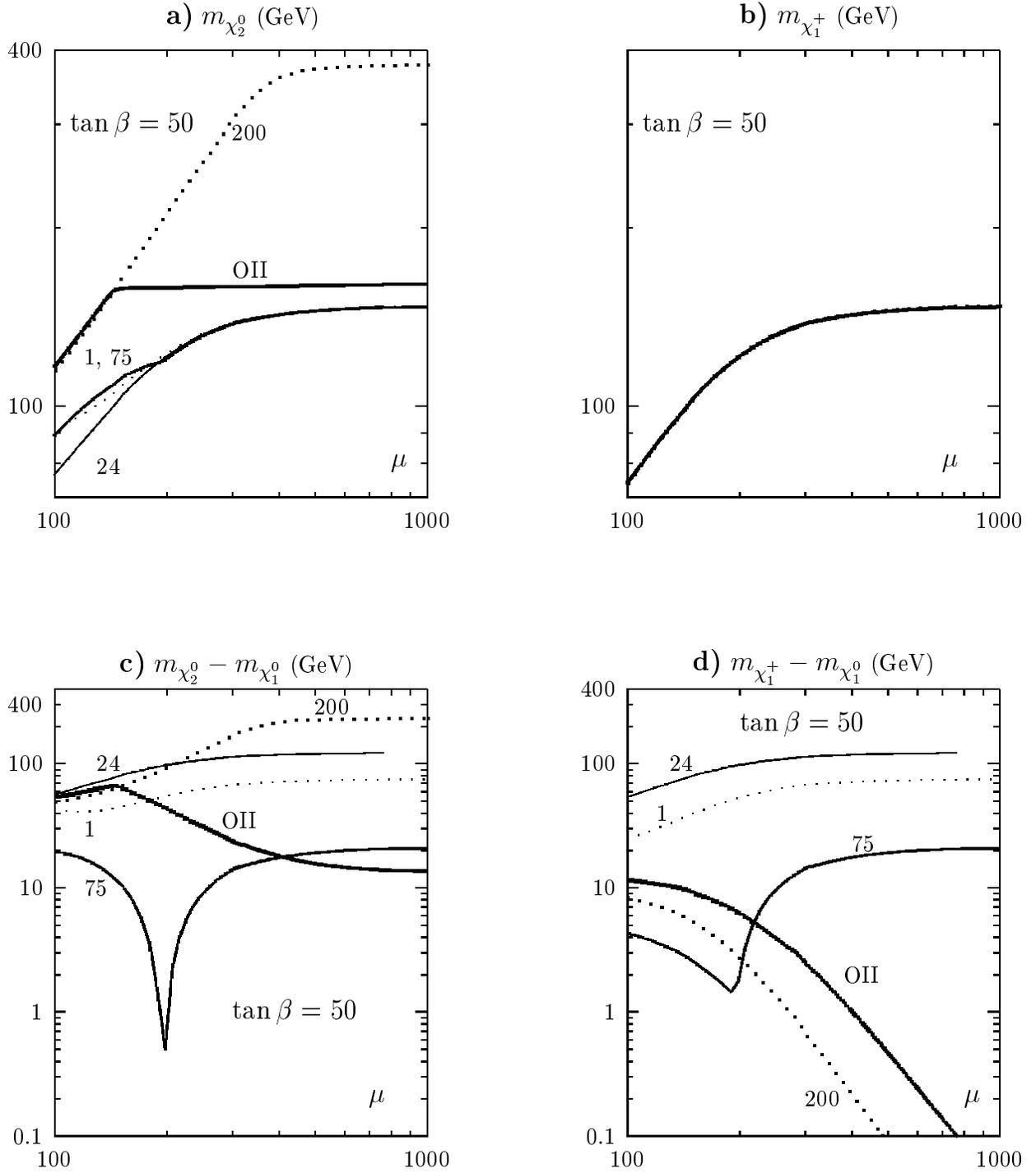,width=22.5cm}}
\vspace*{-9.3cm}
\caption[]{\it The masses of $\chi^0_2$ and $\chi_1^\pm$ and their mass 
differences with the LSP $\chi^0_1$ as a function of $\mu$, for  
$\tb=50$ and $M_2=150$ GeV with the $M_1$ values given in Table 1.} 
\end{figure}

\subsubsection*{2.1.2 The sfermion system} 

The sfermion system is described, in addition to $\tb$ and $\mu$, by three 
parameters for each sfermion species: the left-- and right--handed soft SUSY 
breaking scalar masses $m_{\tilde{f}_L}$ and $m_{\tilde{f}_R}$ and the 
trilinear couplings $A_f$. In the case of the third generation scalar fermions,
the mixing between left-- and right--handed sfermions, which is proportional to 
the mass of the partner fermion, must be included \cite{R12}. 
The sfermion mass matrices read
\beq 
\label{sqmass_matrix}
{\cal M}^2_{\tilde{f}} =
\left(
  \begin{array}{cc} m_f^2 + m_{LL}^2 & m_f \, \tilde{A}_f  \\
                    m_f\, \tilde{A}_f    & m_f^2 + m_{RR}^2 
  \end{array} \right) \ \ {\rm with} 
\begin{array}{l} 
\ m_{LL}^2 =m_{\tilde{f}_L}^2 + (I_3^f - e_f s_W^2)\, M_Z^2\, c_{2\beta} \\\
m_{RR}^2 = m_{\tilde{f}_R}^2 + e_f s_W^2\, M_Z^2\, c_{2\beta} \\\
\ \tilde{A}_f  = A_f - \mu (\tb)^{-2 I_3^f} 
\label{mass-matrix}
\end{array}
\eeq
where $I_3^f$ and $e_f$ are the weak isospin and electric charge of the 
sfermion $\tilde{f}$, and $s_W^2=1-c_W^2 \equiv \sin^2\theta_W$. They are 
diagonalized by $ 2 \times 2$ rotation matrices of angle $\theta_f$,  
which turn the current eigenstates, $\tilde{f}_L$ and $\tilde{f}_R$, into
the mass eigenstates $\tilde{f}_1$ and $\tilde{f}_2$; the mixing angle and 
sfermion masses are then given by 
\beq
\sin 2\theta_f = \frac{2 m_f \tilde{A}_f} { m_{\tilde{f}_1}^2
-m_{\tilde{f}_2}^2 } \ \ , \ \ 
\cos 2\theta_t = \frac{m_{LL}^2 -m_{RR}^2} 
{m_{\tilde{f}_1}^2 -m_{\tilde{f}_2}^2 } \hspace*{0.8cm}  \\
m_{\tilde{f}_{1,2}}^2 = m_f^2 +\frac{1}{2} \left[
m_{LL}^2 +m_{RR}^2 \mp \sqrt{ (m_{LL}^2
-m_{RR}^2 )^2 +4 m_f^2 \tilde{A}_f^2 } \ \right]
\eeq
The mixing is very strong in the stop sector for large values of $\tilde{A}_t$
and makes the lightest $\tilde{t}_1$ much lighter than the other squarks and
possibly even lighter than the top quark itself. For large values of $\tan 
\beta$ and $\mu$, the mixing in the sbottom and stau sectors can be also very 
strong, $\tilde{A}_{b,\tau} \sim - \mu \tb$, leading to lighter $\tilde{b}_1$ 
and $\tilde{\tau}_1$ states. \s

Since the fermion masses provide one of the main inputs for sfermion mixing, it
is important to include the leading radiative corrections to these parameters 
\cite{radmf}, in particular those due to strong interactions. The fermion 
masses which have to be used in the mass matrices eq.~(\ref{mass-matrix}) are 
the masses $\hat{m}_f (Q^2)$, evaluated in the $\overline{\rm DR}$ scheme at 
the scale $Q$ and which, in terms of the pole masses $m_f$, are given by 
\cite{Radcor}:  
\beq 
m_f = \hat{m}_f (Q^2) \, \left(1+ \frac{\Delta m_f}{m_f} \right) 
\eeq
In the case of top quarks, it is sufficient to include the one--loop QCD 
corrections originating from standard gluon exchange (first term) and 
gluino--stop exchange (second term):
\begin{eqnarray}
\frac{\Delta m_t}{m_t} &=& \frac{\alpha_s}{3\pi} \left[3 \log \left( \frac{Q^2}
{m_t^2} \right) +5 \right]  \\
&-& \frac{\alpha_s}{3\pi} \left[ B_1(m_{\tilde{g}}, m_{\tilde{t}_1})+
B_1( m_{\tilde{g}}, m_{\tilde{t}_2}) -s_{2\theta_t} \frac{m_{\tilde{g}}}{m_t}
\bigg(B_0(m_{\tilde{g}}, m_{\tilde{t}_1})-B_0(m_{\tilde{g}}, m_{\tilde{t}_2}) 
\bigg) \right] \non 
\end{eqnarray}
where in terms of $M={\rm max}(m_1,m_2)$, $m={\rm min}(m_1,m_2)$ and $x=
m_2^2/m_1^2$, the two Passarino--Veltman functions \cite{PV} $B_{0,1}(m_1,m_2) 
\equiv B_{0,1}(0,m_1^2,m_2^2)$ simply read in this limit
\begin{eqnarray}
B_0 (m_1,m_2) &=& -\log \left( \frac{M^2}{Q^2} \right) +1 + \frac{m^2}{m^2-M^2}
\log \left( \frac{M^2}{m^2} \right) \non \\
B_1 (m_1,m_2) &=& \frac{1}{2} \left[- \log \left( \frac{M^2}{Q^2} \right) 
+\frac{1}{2}  + \frac{1}{1-x}+ \frac{\log x}{(1-x)^2} - \theta (1-x) \log x
\right]
\end{eqnarray}
In the case of bottom quarks, the first important correction which has to be 
included is the one due to standard QCD corrections and the running from the 
scale $m_b$ to the high scale $Q$. The $\overline{\rm DR}$ $b$--quark mass 
[for the NNLO corrections, we assume that the correction in the 
$\overline{\rm MS}$ and $\overline{\rm DR}$ schemes are the same, since the 
latter is not yet available] is given by \cite{bmass}: 
\beq
\hat{m}_b(Q^2) = \hat{m}_b(m_b^2) \, c[\alpha_s(Q^2)/\pi]\, / \, 
c[\alpha_s(m_b^2)/ \pi] 
\eeq
with  
\beq 
\hspace*{1.3cm}
\hat{m}_b(m_b^2) &=& m_b \left[1+ \frac{5}{3} \frac{\alpha_s(m_b^2)}{\pi}
+ 12.4 \frac{\alpha^2_s(m_b^2)}{\pi^2} \right] \\
c(x)&=&(23x/6)^{12/23} [1+1.175x + 1.5 x^2] \ \ {\rm for}  \ Q^2 < m_t^2 \non \\
c(x)&=&(7x/2)^{4/7} [1+1.398x + 1.8 x^2] \ \ \ \ \ \  {\rm for} \ Q^2 > m_t^2 
\eeq
After this, one has to include the sbottom--gluino and the stop--chargino 
corrections which are the most important ones \cite{Radcor}, in particular for 
large $\tb$ and $\mu$ values:
\begin{eqnarray}
\frac{\Delta m_b}{m_b} &=&-\frac{\alpha_s}{3\pi} \left[ B_1(m_{\tilde{g}}, 
m_{\tilde{b}_1})+B_1( m_{\tilde{g}}, m_{\tilde{b}_2}) -s_{2\theta_b} 
\frac{m_{\tilde{g}}}{m_b} \bigg( B_0(m_{\tilde{g}}, m_{\tilde{b}_1})-
B_0(m_{\tilde{g}}, m_{\tilde{b}_2}) \bigg) \right] \non \\
&-& \frac{\alpha}{8\pi s_W^2} \frac{m_t \mu}{M_W^2 \sin 2\beta} \, s_{2\theta_t}
\, [B_0(\mu, m_{\tilde{t}_1})- B_0( \mu, m_{\tilde{t}_2}) ] \non \\
&-& \frac{\alpha}{4\pi s_W^2} \left[ \frac{M_2\mu \tb}{\mu^2-M_2^2}
\bigg( c^2_{\theta_t} B_0(M_2, m_{\tilde{t}_1})+ s_{\theta_t}^2 B_0(M_2, 
m_{\tilde{t}_2}) \bigg) + (\mu \leftrightarrow M_2) \right]
\end{eqnarray}
For the $\tau$ lepton mass, the only relevant corrections to be included are 
those stemming from chargino--sneutrino loops, and which simply read
\begin{eqnarray}
\frac{\Delta m_\tau}{m_\tau} &=& -\frac{\alpha}{4\pi s_W^2} \, 
\frac{M_2\mu \tb}{\mu^2-M_2^2} \left[ B_0(M_2, m_{\tilde{\nu}_\tau})
-B_0(\mu, m_{\tilde{\nu}_\tau}) \right] 
\end{eqnarray}
The effect of the radiative corrections is shown in Fig.~2 for the case of the
bottom quark and tau lepton masses for $\tb=50$ as a function of $\mu$ for the
various models with and without unification of the gaugino masses at $M_{\rm
GUT}$. The wino mass is fixed to $M_2=150$ GeV and $M_1,M_3 \simeq m_{\tilde g}
$ at the weak scale are given in Table 1. The main correction to the ${\rm
\overline{DR}}$ bottom quark mass, $\hat{m}_b (M_Z^2) \sim 3$ GeV, is due to
the SUSY--QCD corrections from gluino--sbottom loops in the case of large 
values of $\tb$ and $\mu$. This correction is proportional to $\Delta m_b \sim -
(\alpha_s /\pi ) \times  \tb \mu m_{\tilde{g}} / m_{\tilde{b}}^2$ and can
increase or decrease [depending of the sign of $\mu$] the $b$--quark mass by
more than a factor of two.  The effect of the radiative corrections is less
drastic in the case of the $\tau$ mass since the latter are of the order a few
percent. \s

Let us now discuss the dependence of the sfermion masses on the gaugino masses
as well as on the parameters $\mu$ and $\tb$, in models with a universal mass 
$m_0$ for the scalar fermions at the scale $M_{\rm GUT}$, but without the 
gaugino mass unification assumption $m_{1,2,3}=m_{1/2}$. In the case of the 
partners of the light fermions [including $b$--quarks], one can neglect to a 
good approximation the effect of the Yukawa couplings in the one--loop 
Renormalization Group evolution of the soft SUSY breaking scalar masses. With 
the notation of the first generation, one then obtains, when including the 
D--terms, the following expressions \cite{Komine}:
\begin{eqnarray}
m^2_{\tilde{u}_L} &=& m_0^2 + 5.8 m^2_3+0.47 m^2_2 + 4.2 \times 10^{-3} m^2_1 
+ 0.35 M_Z^2 \cos2\beta \non \\
m^2_{\tilde{d}_L} &=& m_0^2 + 5.8 m^2_3+0.47 m^2_2 + 4.2 \times 10^{-3} m^2_1 
-0.42 M_Z^2 \cos2\beta \non \\
m^2_{\tilde{u}_R} &=& m_0^2 + 5.8 m^2_3+ 6.6 \times 10^{-2} m^2_1 
+ 0.16 M_Z^2 \cos2\beta \non \\
m^2_{\tilde{d}_R} &=& m_0^2 + 5.8 m^2_3+ 1.7 \times 10^{-2} m^2_1 
-0.08 M_Z^2 \cos2\beta \non \\
m^2_{\tilde{\nu}_L} &=& m_0^2  +0.47 m^2_2 + 3.7 \times 10^{-2} m^2_1 
+ 0.50 M_Z^2 \cos2\beta \non \\
m^2_{\tilde{e}_L} &=& m_0^2 + 0.47 m^2_2 + 3.7 \times 10^{-2} m^2_1 
 -0.27 M_Z^2 \cos2\beta \non \\
m^2_{\tilde{e}_R} &=& m_0^2 +0.15 m^2_{1} -0.23 M_Z^2 \cos2\beta 
\label{smass}
\end{eqnarray}
One has then, in the case of sbottoms and staus, to include the mixing since in
this case, large enough off--diagonal elements of the mass matrices are
obtained for large $\mu$ and $\tb$ values [the effect of the trilinear
couplings plays only a marginal role]. \s

The squark masses are governed by the parameter $m_3$, while the slepton masses
are governed by the parameter $m_2$, and to a lesser extent $m_1$.  Figs.~3a-b
show the variation of the soft parameters $m_{\tilde{b}_1}$ (a) and $m_{\tilde{
\tau}_1}$ (b) as a function of $M_2$ for $\tb=50$ and $m_0=300$ GeV. As
can be seen, depending on the models, the squark and slepton masses can be
different for different models.  In Fig.~3c, the masses $m_{\tilde{d}_R},
m_{\tilde{e}_R}$ and $m_{\tilde{b}_1}, m_{\tilde{\tau}_1}$ are shown as a
function of $\mu$ for $\tb=50$; we have used the previous equations and fixed
$m_0=300$ GeV and $m_1=m_2=m_3=m_{1/2}= 120$ GeV, i.e. as in the mSUGRA--type
scenario. While for small values of $\mu$, and hence small off--diagonal
elements in the $\tilde{b}$ and $\tilde{\tau}$ mass matrices, $\tilde{d}_R,
\tilde{b}_1$ and $\tilde{e}_R, \tilde{\tau}_1$ are almost degenerate in mass,
the mass splitting increases with increasing $\mu$ reaching a substantial
amount for $\mu \geq m_0$.  

\begin{figure}[htbp]
\vspace*{-4.5cm}
\hspace*{-2.3cm}
\mbox{\psfig{figure=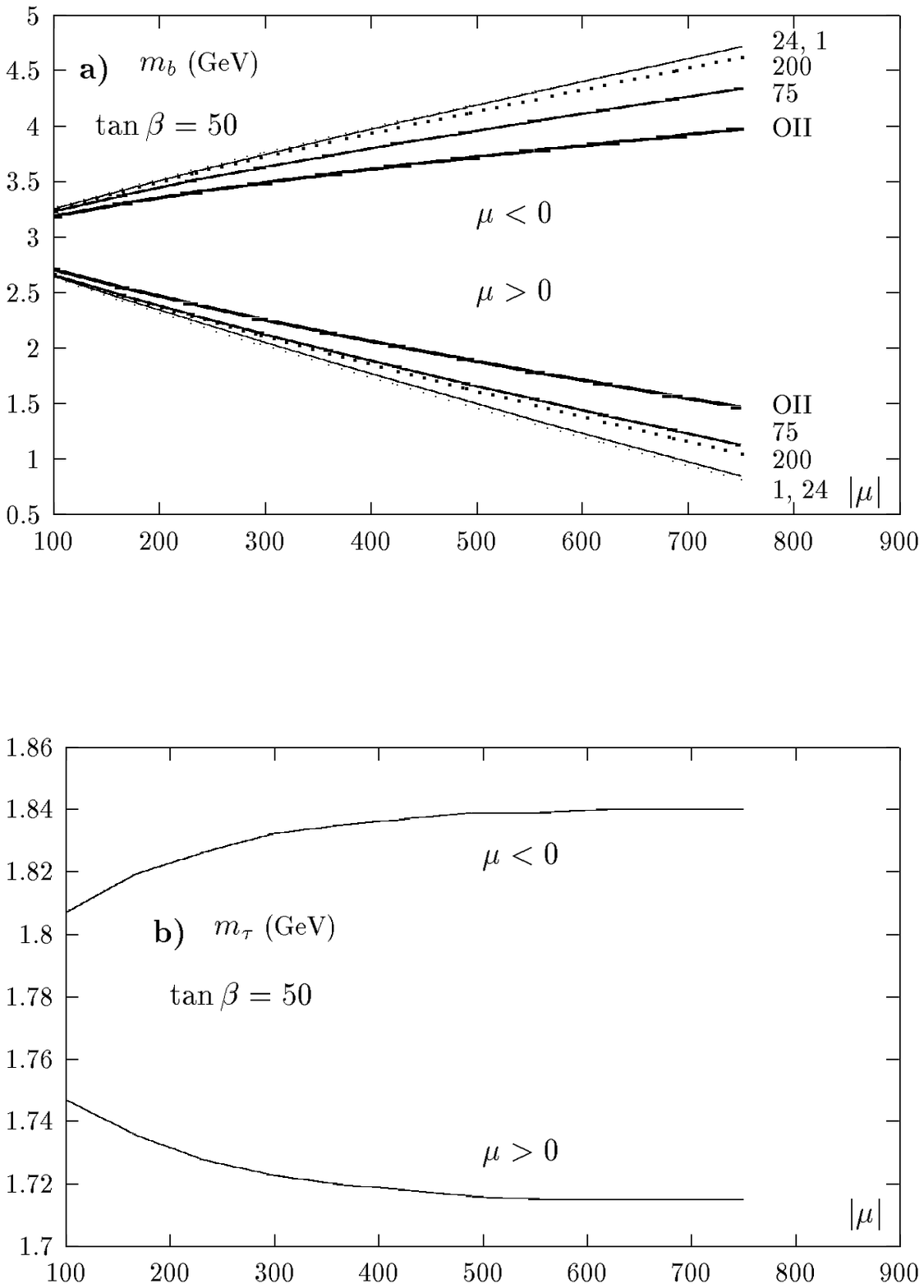,width=22.5cm}}
\vspace*{-9.3cm}
\caption[]{\it The $b$ quark and $\tau$ lepton masses, including the radiative
corrections, as a function of $\mu$, for $\tb=50$ and $M_2=150$ GeV in the 
various models of Table 1.} 
\end{figure}

\begin{figure}[htbp]
\vspace*{-4.5cm}
\hspace*{-2.3cm}
\mbox{\psfig{figure=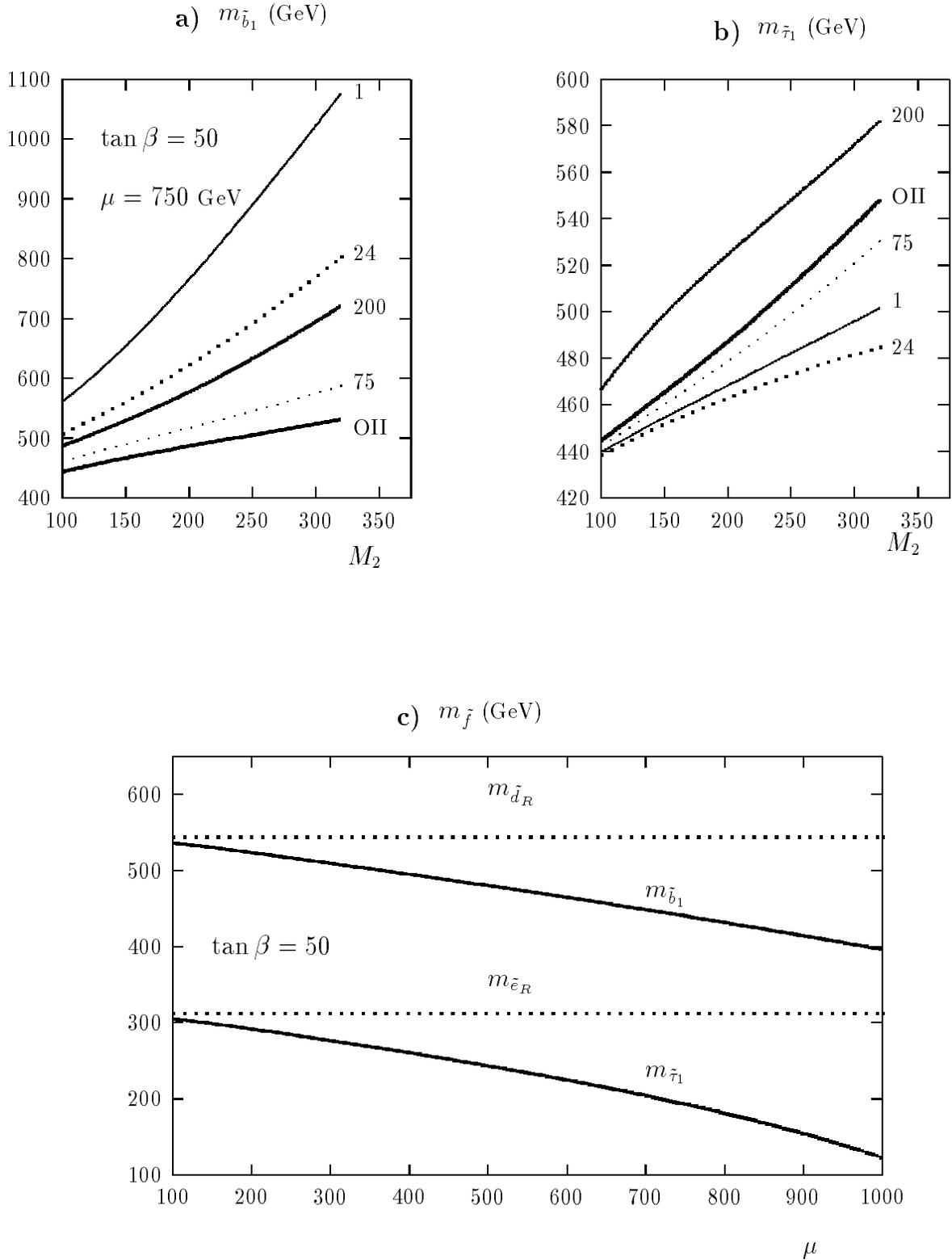,width=22cm}}
\vspace*{-8.cm}
\caption[]{\it The masses of lightest sbottom (a) and tau slepton (b) as a 
function of $M_{2}$ for $m_0=300$ GeV and $\mu=750$ GeV in the models of Table 
1. The $\tilde{b}_1, \tilde{d}_R$ and $\tilde{\tau}_1, \tilde{e}_R$ masses as 
a function of $\mu$, for $\tb=50$ and $M_2=150$ GeV in model ${\bf 1}$ (c).} 
\end{figure}

\subsubsection*{2.2.3 The Higgs sector} 

The MSSM includes two iso--doublets of Higgs fields, which after spontaneous
symmetry breaking, give rise to a quintet of physical Higgs boson states: 
$h$, $H$, $A$, $H^\pm$ \cite{R13}.  While an upper bound of about 130~GeV can 
be derived on the mass of the light CP--even neutral Higgs boson $h$ \cite{R14},
the heavy CP--even and CP--odd neutral Higgs bosons $H$, $A$, and the charged 
Higgs bosons $H^\pm$ may have masses of the order of the electroweak symmetry 
scale $v$ up to about 1~TeV.  This extended Higgs system can be described by 
two parameters at the tree level: tan$\beta$ and one mass parameter which is 
generally identified with the pseudoscalar mass $M_A$.
The Higgs mass parameters and the couplings are affected by top and stop loop
radiative corrections \cite{R14}, which in the leading approximation are 
parameterized by
\beq
\epsilon \approx \frac{3 G_F m_t^4}{\sqrt{2} \pi^2 \sin^2 \beta} 
\log \frac{\tilde{m}^2}{m_t^2} 
\label{epsilon}
\eeq
where the scale of supersymmetry breaking is characterized by a common
squark--mass value $\tilde{m}$. The next--to--leading order QCD corrections
can be included by using the running top quark mass in the $\overline{\rm MS}$
scheme. Stop mixing effects can be accounted for by 
shifting $\tilde{m}^2$ in eq.~(\ref{epsilon}) by 
the amount [$\tilde{A_t} = A_t - \mu \cot\beta$]
\begin{eqnarray}
\tilde{m}^2 \to \tilde{m}^2 + \Delta \tilde{m}^2 &:& \Delta \tilde{m}^2 
= \tilde{A}_t^2 [1-\tilde{A}_t^2/(12\tilde{m}^2)] 
\end{eqnarray}
The neutral CP--even and charged Higgs boson masses and the mixing angle 
$\alpha$ in the  neutral sector, when expressed in terms of $M_A$ 
and tan $\beta$, are given in this approximation by 
\beq
M_{h,H}^2 \!\!&=&\!\! \frac{1}{2} \left[ M_A^2+M_Z^2+\epsilon \mp
\sqrt{ (M_A^2+M_Z^2+\epsilon)^2- 4M_A^2 M_Z^2 c_{2\beta}^2
   - 4\epsilon( M_A^2 s^2_\beta + M_Z^2 c^2_\beta) } \right] \non \\
M_{H^\pm}^2 \!\!&=&\!\! M_W^2 +M_A^2 \non \\
\tan 2\alpha \!\!&=&\!\! \tan 2\beta
 \frac{M_A^2 + M_Z^2}{M_A^2 - M_Z^2 +\epsilon/c_{2\beta} } \qquad
\mbox{with} \qquad  - \frac{\pi}{2} \leq \alpha \leq 0
\label{mass}
\eeq
In the decoupling limit, $M_A \gg M_Z$, the $A,H,H^\pm$ bosons become 
degenerate in mass $M_A \simeq M_H \simeq M_{H^\pm}$ while the lightest $h$
boson reaches its maximal mass value $M_h^2 \sim M_Z^2 +\epsilon$; the angle 
$\alpha$ approaches the value $\alpha \to \beta- \pi/2$. The couplings of
the $h$ particle to fermions and gauge bosons are then SM--like, while the
couplings of the $H,A,H^\pm$ bosons  to down (up)--type fermions are 
(inversely) proportional to $\tb$. \s

In the present analysis, we will use the full renormalization--group improved 
radiative corrections to the Higgs sector given in Ref.~\cite{SUBH}. 
We will often denote the Higgs bosons by $H_k$ with $k=1,2,3,4$, corresponding 
to $H,h, A$ and $H^\pm$, respectively. 

\subsection*{2.2 Couplings}

In this subsection, we list the various couplings \cite{Haber-Kane,R9,R10} 
which will be needed in our analysis. All the couplings are normalized to the 
electric charge $e$. \s

\nn $\bullet$ The couplings of the charginos and neutralinos to the weak gauge 
bosons $W^\pm, Z$: 
\beq
G^{L,R}_{\chi^0_i \chi^+_j W^+} = G^{L,R}_{ijW} & {\rm with} & 
\begin{array}{l} 
G^L_{ijW} =  \frac{1}{\sqrt{2}s_W} [-Z_{i4} V_{j2}+\sqrt{2}Z_{i2} V_{j1}] \\
G^R_{ijW} =  \frac{1}{\sqrt{2}s_W} [Z_{i3} U_{j2}+ \sqrt{2} Z_{i2} U_{j1}] 
\end{array}  
\eeq
\beq
G^{L,R}_{\chi^-_i \chi^+_j Z} = G^{L,R}_{ijZ} & {\rm with} & 
\begin{array}{l} 
G^L_{ijZ} = \frac{1}{c_W s_W} \left[- \frac{1}{2} V_{i2} V_{j2} - 
V_{i1} V_{j1} +\delta_{ij}s_W^2 \right]  \\
G^R_{ijZ} = \frac{1}{c_W s_W} \left[- \frac{1}{2} U_{i2} U_{j2} - 
U_{i1} U_{j1} +\delta_{ij}s_W^2 \right] 
\end{array} 
\eeq
\beq
G^{L,R}_{\chi^0_i \chi^0_j Z} = G^{L,R}_{ijZ} & {\rm with} &
\begin{array}{l} 
G^L_{ijZ} = - \frac{1}{2s_Wc_W} [Z_{i3} Z_{j3} - Z_{i4} Z_{j4}]  \\
G^R_{ijZ} = + \frac{1}{2s_Wc_W} [Z_{i3} Z_{j3} - Z_{i4} Z_{j4} ] 
\end{array} 
\eeq

\nn $\bullet$ The couplings of charginos and neutralinos to the Higgs bosons: 
\beq
G^{L,R}_{\chi^0_i \chi^+_j H^+} = G^{L,R}_{ij4} & {\rm with} &  
\begin{array}{l} 
G^L_{ij4} = \frac{c_\beta}{s_W} \left[ Z_{j4} V_{i1} + \frac{1}{\sqrt{2}} 
\left( Z_{j2} + \tan \theta_W Z_{j1} \right) V_{i2} \right] \non \\
G^R_{ij4} = \frac{s_\beta}{s_W} \left[ Z_{j3} U_{i1} - \frac{1}{\sqrt{2}}
\left( Z_{j2} + \tan \theta_W Z_{j1} \right) U_{i2} \right]
\end{array} 
\eeq
\beq
G^{L,R}_{\chi^-_i \chi^+_j H^0_k} = G^{L,R}_{ijk} & {\rm with} & 
\begin{array}{l} 
G^L_{ijk}= \frac{1}{\sqrt{2}s_W} \left[ e_k V_{j1}U_{i2}-d_k V_{j2}U_{i1}
\right] \\
G^R_{ijk}= \frac{1}{\sqrt{2}s_W} \left[ e_k V_{i1}U_{j2}-d_k V_{i2}U_{j1}
\right] \epsilon_k 
\end{array} 
\eeq
\beq
G^{L,R}_{\chi^0_i \chi^0_j H_k} = G^{L,R}_{ijk} & {\rm with} & 
\begin{array}{l} 
G^L_{ijk} = \frac{1}{2 s_W} \left( Z_{j2}- \tan\theta_W Z_{j1} \right) 
\left(e_k Z_{i3} + d_kZ_{i4} \right) \ + \ i \leftrightarrow j
 \\
G^R_{ijk} = \frac{1}{2 s_W}  \left( Z_{j2}- \tan\theta_W Z_{j1} 
\right) \left(e_k Z_{i3} + d_kZ_{i4} \right) \epsilon_k \ + \ i 
\leftrightarrow j
\end{array} 
\eeq
where $\epsilon_{1,2}=- \epsilon_3 =1$ and the coefficients $e_k$ and $d_k$ 
read
\begin{eqnarray}
e_1/d_1=c_\alpha/  -s_\alpha \ , \
e_2/d_2=-s_\alpha / -c_\alpha \ ,  \
e_3/d_3=-s_\beta / c_\beta
\end{eqnarray}

\nn $\bullet$ For the couplings between neutralinos, fermions and sfermions, 
$\tilde f_i-f-\chi_j^0$, one has: 
\begin{eqnarray}
\left\{ \begin{array}{c} a^{\tilde f}_{j1} \\  a^{\tilde f}_{j2} \end{array} 
\right\}
        &=&   -\frac{m_f r_f}{\sqrt{2} M_W s_W}
\left\{ \begin{array}{c} \st{f} \\ \ct{f} \end{array} 
\right\}
         - e^f_{Lj}\, \left\{ \begin{array}{c} \ct{f} \\ -\st{f} \end{array} 
\right\} \nonumber \\
\left\{ \begin{array}{c} b^{\tilde f}_{j1} \\  b^{\tilde f}_{j2} \end{array} 
\right\}
        &=&  -\frac{m_f r_f}{\sqrt{2} M_W s_W }
\left\{ \begin{array}{c} \ct{f} \\ -\st{f} \end{array} 
\right\}
         - e^f_{Rj}\, \left\{ \begin{array}{c} \st{f} \\ \ct{f} \end{array} 
\right\}
\end{eqnarray}
with $r_u= Z_{j4}/ \sin \beta$ and $r_d=Z_{j3}/\cos \beta$ for up and 
down--type fermions, and  
\beq
e^f_{Lj} & = & \sqrt{2}\left[ e_f \; Z_{j1}'
              + \left(I_f^3 - e_f \, s_W^2 \right)
                 \frac{1}{c_W s_W}\;Z_{j2}' \right]  \nonumber \\ 
e^f_{Rj} & = &-\sqrt{2} \, e_f \left[ Z_{j1}'
              - \frac{s_W}{c_W} \; Z_{j2}' \right]  
\eeq

\nn $\bullet$ For the couplings between charginos, fermions and sfermions, 
$\tilde f_i-f'-\chi_j^+$, one has for up--type and down--type sfermions: 
\beq
\left\{ \begin{array}{c} a^{\tilde u}_{j1} \\ a^{\tilde u}_{j2} \end{array} 
\right\} & = &
   \frac{V_{j1}}{s_W} \,
\left\{ \begin{array}{c} -\ct{u} \\ \st{u} \end{array} \right\}
 + \frac{m_u\,V_{j2}}{\sqrt{2}\,M_W s_W\,s_\beta}
\left\{ \begin{array}{c} \st{u} \\ \ct{u} \end{array} \right\} 
\non \\
\left\{ \begin{array}{c} b^{\tilde u}_{j1} \\ b^{\tilde u}_{j2} \end{array} 
\right\} & = &
    \frac{m_d\,U_{j2}}{\sqrt{2}\,M_W s_W \, c_\beta}
\left\{ \begin{array}{c} \ct{u} \\ -\st{u} \end{array} \right\}
\eeq
\beq
\left\{ \begin{array}{c} a^{\tilde d}_{j1} \\ a^{\tilde d}_{j2}
\end{array} \right\} & = &
   \frac{U_{j1}}{s_W} \,
\left\{ \begin{array}{c} -\ct{d} \\ \st{d} \end{array} \right\}
 + \frac{m_d\,U_{j2}}{\sqrt{2}\,M_W s_W \,c_\beta}
\left\{ \begin{array}{c} \st{d} \\ \ct{d} \end{array} \right\} 
\non \\
\left\{ \begin{array}{c} b^{\tilde d}_{j1} \\ b^{\tilde d}_{j2} \end{array} 
\right\} & = &
   \frac{m_u\,V_{j2}}{\sqrt{2} \,M_W s_W \,s_\beta}
\left\{ \begin{array}{c} \ct{d} \\ -\st{d} \end{array} \right\}
\eeq
\smallskip

\nn $\bullet$ Finally, the couplings of the $W,Z$ gauge bosons and the four 
Higgs bosons $H_k=H,h,A,H^\pm$ with $k=1,..,4$ to fermions are: 
\beq
v_{Z}^f= \frac{2I_f^3- 4 e_f s_W^2}{4c_Ws_W} \ , \ 
a_{Z}^f= \frac{2I_f^3}{4c_Ws_W} \ , \ v_{W}^f =a_{W}^f = \frac{1}{2\sqrt{2}s_W}
\eeq
\beq
v_{1}^f =\frac{m_f r_2^f}{2s_W M_W}   \ , \ a_1^f =0 \ , \ 
v_{2}^f =\frac{m_f r_1^f }{2s_W M_W}   \ , \ a_2^f =0 \ , \   
v_3^f = 0 \ ,\ a_{3}^f =\frac{-m_f (\tb)^{-2I_f^3}}{2s_W M_W} \hspace*{1.1cm}
\eeq
\beq
v_{4}^f= -\frac{m_d \tb + m_u {\rm cot}\beta}{2\sqrt{2} s_W M_W} \ , \
a_{4}^f= \frac{m_d \tb - m_u {\rm cot} \beta}{2\sqrt{2} s_W M_W} 
\eeq
with the coefficients $r^f_{1,2}$ as 
\beq
r_1^u = s_\alpha/s_\beta \ \  , \ \ 
r_2^u = c_\alpha/s_\beta \ \ , \ \
r_1^d = c_\alpha / c_\beta \ \ , \ \ 
r_2^d = -s_\alpha /c_\beta 
\eeq

\newpage

\section*{3. Three--body decays} 
\setcounter{equation}{0}
\renewcommand{\theequation}{3.\arabic{equation}}

In this section we give the complete analytical expressions of the partial 
widths of the three--body decays of charginos and neutralinos into 
a neutralino and two fermions, that we will denote to be general by $u$ 
and $\bar{d}$ [although they can be the same] 
\beq 
\label{decay}
\chi_i \, \rightarrow \, \chi_j^0  \ u \ \bar{d} 
\eeq
We will not assume that the final neutralino is the LSP $\chi_1^0$, but any of
the neutralinos $\chi_j^0$ to cover also the possibility of cascade decays.  As
shown in Fig.~4, these decays proceed through gauge boson exchange [$V=W$ and
$Z$ for $\chi_i^+$ and $\chi_i^0$ decays, respectively], Higgs boson exchange
[$H_k=H^+$ for $\chi_i^+$ decays and $H_k=H,h,A$ with $k=1,2,3$ for $\chi_i^0$
decays] and sfermion exchange in the $t$-- and $u$--channels [the flavor is
fixed  by the sfermion--fermion and final neutralino vertex].  For gluino
decays \cite{gluino,N1}, only the channels with $u$ and $t$--channel squark
exchange will be present; the partial widths can be straightforwardly derived
from those of the neutralino decays, with the appropriate change of the
couplings. Note that for the treatment of the Majorana nature of the 
initial state, we use the rules given in Ref.~\cite{denner}. \bigskip

\begin{picture}(1000,170)(10,0)
\Text(20,130)[]{$\chi_i$}
\ArrowLine(10,120)(40,120)
\ArrowLine(40,120)(70,150)
\Text(77,150)[]{$\chi_j^0$}
\Photon(40,120)(60,90){4}{8}
\Text(35,100)[]{$V$}
\ArrowLine(60,90)(90,120)
\Text(95,120)[]{$u$}
\ArrowLine(80,65)(60,90)
\Text(90,65)[]{$\bar{d}$}
\Text(140,130)[]{$\chi_i$}
\ArrowLine(130,120)(160,120)
\ArrowLine(160,120)(190,150)
\Text(197,150)[]{$\chi_j^0$}
\DashArrowLine(160,120)(180,90){4}
\Text(155,100)[]{$\Phi$}
\ArrowLine(180,90)(210,120)
\Text(215,120)[]{$u$}
\ArrowLine(200,65)(180,90)
\Text(210,65)[]{$\bar{d}$}
\Text(270,130)[]{$\chi_i$}
\ArrowLine(250,120)(280,120)
\ArrowLine(280,120)(310,150)
\Text(317,150)[]{$u$}
\DashArrowLine(280,120)(300,90){4}{}
\Text(275,100)[]{$\tilde{f}$}
\ArrowLine(300,90)(330,120)
\Text(340,120)[]{$\bar{d}$}
\ArrowLine(320,65)(300,90)
\Text(330,65)[]{$\chi_j^0$}
\Text(390,130)[]{$\chi_i$}
\ArrowLine(370,120)(400,120)
\ArrowLine(400,120)(430,150)
\Text(437,150)[]{$\bar{d}$}
\DashArrowLine(400,120)(420,90){4}{}
\Text(395,100)[]{$\tilde{f}$}
\ArrowLine(420,90)(450,120)
\Text(455,120)[]{$u$}
\ArrowLine(440,65)(420,90)
\Text(450,65)[]{$\chi_j^0$}
\end{picture}
\vspace*{-1.5cm}

\nn {\it Figure 4: The Feynman diagrams contributing to the three--body decays 
of charginos and neutralinos into the LSP and two fermions.}  
\vspace*{4mm} 
\setcounter{figure}{4}

In this section, we will simply give the complete
analytical expressions for the (unpolarized) Dalitz plot density in terms of
the energies of two final fermions, and for the fully integrated partial widths
for vanishing fermion masses\footnote{In mSUGRA--type models, this 
approximation is very
good for all light fermion final states, including $b$--quarks and $\tau$
leptons, since  $\chi_2^0$ and $\chi_1^+$ are expected to have masses larger
than ${\cal O}(100$ GeV). The approximation would be bad for top quark final
states; however, if the three--body decays $\chi_2^0 \to \chi_1^0 t \bar{t}$
and $\chi_1^+ \to \chi_1^0 t \bar{b}$ are kinematically allowed, they will not
play a major role since the charginos and neutralinos will have enough phase
space to decay first into the two--body channels $\chi_2^0 \to \chi_1^0 Z,\;
\chi_1^0 h$ [and possibly $\chi_1^0 H$ and $\chi_1^0 A$] and $\chi_1^+ \to 
\chi_1^0 W$ [and possibly $\chi_1^0 H^+$], which will be largely dominating.}.
[In the most complete analysis of these decays available in the
literature up to now, Ref.~\cite{R4a}, the fully integrated partial widths have
not been derived: one integral has been left--out and performed numerically.]
The formulae for the general case with non--vanishing values for the masses of
the final standard fermions [to be able to describe more accurately the cases
of chargino decays into $\tau \nu$ as well as neutralino decays into $b\bar{b}
$ and $\tau^+ \tau^-$ final states and to treat the case of the top quark] and
where the polarization of the initial gauginos are taken into account, are
given in the Appendix. 

\subsection*{3.1 The Dalitz densities for the three--body decays}

The Dalitz density of the decay mode eq.~(\ref{decay}) is given in terms 
of the reduced energies of the two final state fermions 
\begin{eqnarray} 
x_1=2E_u/m_{\chi_i} \ \ , \ \ x_2= 2E_d/m_{\chi_i} 
\ \ , \ \ x_3= 2E_{\chi_j}/m_{\chi_i}=2-x_1-x_2
\end{eqnarray}
but we will also use the simplifying notation:
\beq
y_1=1-x_1 - \mu_\chi \ \ ,  \ \
y_2=1-x_2 - \mu_\chi \ \ ,  \ \
y_3=1-x_3 + \mu_\chi 
\eeq
with the reduced masses  $\mu^2_X= M_X^2/m^2_{\chi_i}$ [for the final state 
neutralino we 
drop the index, i.e. $\mu_\chi= m^2_{\chi_j^0}/ m^2_{\chi_i}$]. Neglecting 
the masses of the final fermions [but not in the couplings] and the widths
of the exchanged (s)particles,  the Dalitz density is given by: 
\beq
\frac{ \dx \Gamma_{\chi_i} } {\dx x_1 \dx x_2}  &=& \frac{e^4 m_{\chi_i}} 
{64(2\pi)^3} \, N_c \\ && \bigg[ \dx \Gamma_V + \dx \Gamma_{\tilde{u}}
+ \dx \Gamma_{\tilde{d}} + \dx \Gamma_{\Phi} +\dx \Gamma_{H_1 H_2} + \dx 
\Gamma_{V \tilde{u}} +  
\dx \Gamma_{V \tilde{d}} + \dx \Gamma_{\tilde{u} \tilde{d}} + \dx \Gamma_{\Phi 
\tilde{u}} + \dx \Gamma_{\Phi \tilde{d}}  \bigg] \non 
\eeq
where $N_c$ is the color factor [$N_c=3(1)$ for final state quarks (leptons)]
and the d$\Gamma$'s correspond, respectively, to the separate contributions of
the square of the gauge boson, $\tilde{u}$, $\tilde{d}$ and Higgs exchanges and
the $V\tilde{u}$, $V\tilde{d}$, $\tilde{u} \tilde{d}$,  $\Phi \tilde{u}$, 
$\Phi \tilde{d}$ and $H_1 H_2$ interferences. \s

The various contributions, in terms of the couplings given in section 2.2, 
read: 
\beq
\dx \Gamma_V &=& \frac{4}{(y_3-\mu_V)^2} \bigg\{   
\bigg[(v_V^f-a_V^f)^2 (G_{jiV}^L)^2 +(v_V^f+a_V^f)^2
(G_{jiV}^R)^2 \bigg]  x_1 y_1 +\bigg[ (v_V^f-a_V^f)^2 
(G_{jiV}^R)^2 \non \\
&& \hspace*{2cm} +(v_V^f+a_V^f)^2 (G_{jiV}^L)^2 \bigg] x_2 y_2 
- 4 [(v_V^f)^2+(a_V^f)^2] G_{jiV}^L G_{jiV}^R \sqrt{\mu_\chi} y_3 \bigg\}  
\eeq
\beq
\dx \Gamma_{\tilde{d}} &=& \sum_{k,l=1}^2
\frac{x_1 y_1}{(1-x_1-\mu_{\tilde{d}_k})(1-x_1-\mu_{\tilde{d}_l})} 
\big( a^d_{ik} a^d_{il} + b^d_{ik} b^d_{il} \big) 
\big( a^d_{jk} a^d_{jl} + b^d_{jk} b^d_{jl} \big) 
\label{amp1dd}
\eeq
\beq
\dx \Gamma_{\tilde{u}} &=& \sum_{k,l=1}^2 \frac{x_2y_2}
{(1-x_2-\mu_{\tilde{u}_k})(1-x_2-\mu_{\tilde{u}_l})} 
\big( a^u_{ik} a^u_{il} + b^u_{ik} b^u_{il} \big) 
\big( a^u_{jk} a^u_{jl} + b^u_{jk} b^u_{jl} \big) 
\label{amp1uu}
\eeq
\beq
\dx \Gamma_{\Phi} = 2 \sum_{k} \frac{ y_3\left[(v^f_k)^2+(a^f_k)^2\right]}
{ (y_3-\mu_{H_k})^2} \bigg[ \bigg( (G_{ijk}^L)^2 + (G_{ijk}^R)^2 \bigg)x_3 
+ 4 \sqrt{\mu_\chi} (G_{ijk}^L G_{ijk}^R) \bigg] 
\eeq
\beq 
\dx \Gamma_{H_1 H_2}= 
\frac{ 4 y_3 v^f_1 v^f_2}{ (y_3-\mu_{H_1})(y_3-\mu_{H_2})} 
\bigg[ \bigg( G_{ij1}^L G_{ij2}^L +  G_{ij1}^R G_{ij2}^R  \bigg)x_3 + 
2 \sqrt{\mu_\chi} (G_{ij1}^L G_{ij2}^R + G_{ij2}^L G_{ij1}^R) \bigg] 
\eeq
\beq
\dx \Gamma_{V \tilde{d}} &=& -4\sum_{k=1}^2\Bigg\{ \frac{\left
[a^d_{ik} a^d_{jk} 
G_{jiV}^R(v^d_V+a^d_V) +b^d_{ik} b^d_{jk} G_{jiV}^L(v^d_V-a^d_V) \right]
x_1y_1 } {(y_3- \mu_V)(1-x_1-\mu_{\tilde{d}_k})} \non \\
&-&\frac{\sqrt{\mu_\chi} \left[a^d_{ik} a^d_{jk} G_{jiV}^L (v^d_V+a^d_V)
+b^d_{ik} b^d_{jk} G_{jiV}^R (v^d_V-a^d_V)\right]y_3} {(y_3- \mu_V)(1-x_1-
\mu_{\tilde{d}_k})}  \Bigg\} \\
\dx \Gamma_{V \tilde{u}} &=& 4\sum_{k=1}^2\Bigg\{ \frac{\left[a^u_{ik} a^u_{jk} 
G_{jiV}^L(v^u_V+a^u_V) + b^u_{ik} b^u_{jk} G_{jiV}^R(v^u_V-a^u_V)\right]
x_2y_2 }{(y_3- \mu_V)(1-x_2-\mu_{\tilde{u}_k})} \non \\
&-&\frac{\sqrt{\mu_\chi} \left[a^u_{ik} a^u_{jk} G_{jiV}^R (v^u_V+a^u_V)
+b^u_{ik} b^u_{jk} G_{jiV}^L (v^u_V-a^u_V)\right]y_3} {(y_3- \mu_V)
(1-x_2-\mu_{\tilde{u}_k})}  \Bigg\}
\eeq
\beq
\dx \Gamma_{\tilde{u} \tilde{d}} &=&  \sum_{k,l=1}^2 \left\{
\frac{(a^u_{jk}a^d_{il} 
b^u_{ik} b^d_{jl} + a^u_{ik} a^d_{jl} b^u_{jk} b^d_{il}) (-x_1 y_1 - x_2y_2 + 
x_3 y_3)}{(1-x_2 -\mu_{\tilde{u}_k})(1-x_1- 
\mu_{\tilde{d}_l})}
\right. \non
\\ 
&+& \left. \frac{2(a^u_{ik} a^u_{jk} a^d_{il} a^d_{jl} 
+ b^u_{ik} b^u_{jk} b^d_{il} b^d_{jl})
\sqrt{\mu_{\chi}}y_3}{(1-x_2 -\mu_{\tilde{u}_k})(1-x_1- 
\mu_{\tilde{d}_l})} \right\}
\label{amp1ud}
\eeq
\beq
\dx \Gamma_{\Phi \tilde{d}}= -\sum_{k,l}\Bigg\{
\frac{(v_k^d-a^d_k)a^d_{il}b^d_{jl}
\left( G^R_{ijk}(x_1y_1-x_2y_2+x_3y_3) + 2G^L_{ijk} \sqrt{\mu_{\chi}}y_3
\right)}{(y_3- \mu_k)(1-x_1-\mu_{\tilde{d}_l})} \non \\
+\frac{(a_k^d+v^d_k)b^d_{il}a^d_{jl}\left( G^L_{ijk}(x_1y_1-x_2y_2+x_3y_3)+2
G^R_{ijk}\sqrt{\mu_{\chi}}y_3 \right)}{(y_3- \mu_k)(1-x_1-\mu_{\tilde{d}_l})}
\Bigg\} \\
\dx \Gamma_{\Phi \tilde{u}}= \sum_{k,l} \Bigg\{\frac{(v_k^u-a^u_k)b^u_{il}
a^u_{jl}\left( G^R_{ijk}(x_1y_1-x_2y_2-x_3y_3) - 2G^L_{ijk} \sqrt{\mu_{\chi}}y_3
\right)}{(y_3- \mu_k)(1-x_2-\mu_{\tilde{u}_l})} \non \\
+\frac{(a_k^u+v^u_k)a^u_{il}b^u_{jl} \left( G^L_{ijk}(x_1y_1-x_2y_2-x_3y_3) 
- 2G^R_{ijk} \sqrt{\mu_{\chi}}y_3 \right)}{(y_3- \mu_k)(1-x_2-
\mu_{\tilde{u}_l})} \Bigg\}
\eeq

A few remarks need to be made at this stage: 

\begin{itemize} 

\item In the expressions of the couplings, the indices $i$ and $j$ refer always 
to the decaying chargino or neutralino and the final state neutralino,
respectively. 

\item For the Higgs boson exchange contributions, in the case of chargino 
decays, only the exchange of the charged 
Higgs boson is present and in ${\rm d} \Gamma_\Phi$ one has $k=4$ only. In the
case of neutralino decays, the three neutral Higgs bosons will contribute
and $k$ in the sum $\sum_k$ of  ${\rm d} \Gamma_\Phi$ runs from $k=1$ to $3$.
In addition, there is an extra term, ${\rm d} \Gamma_{H_1 H_2}$, due to the 
interference between the exchange of the two CP--even Higgs bosons $h$ and 
$H$. Note also that in this case, there is a difference between the 
contributions of the CP--even (and the charged) and CP--odd Higgs bosons which 
appears in the terms $\epsilon_{1,2,4}=1$ and $\epsilon_3=-1$ in the couplings.
 
\item For massless final state fermions, there is no interference between the 
vector boson and Higgs boson contributions. In the Appendix, where the fermion
mass dependence will be included, interference terms between the Higgs bosons
and the vector bosons, which are proportional to the fermion masses,  will 
be shown explicitly.   

\item In the sfermion exchange diagrams, there is a relative minus sign
between the amplitudes of the $u$ and $t$ channels, due to Wick's theorem.   
This leads to  ${\rm d} \Gamma_{V \tilde{u}}$ and ${\rm d}\Gamma_{V \tilde{d}}$
contributions which are anti--symmetric in the interchange of $x_1$ and $x_2$. 
In the case of ${\rm d}\Gamma_{\Phi\tilde{u}}$ and ${\rm d} \Gamma_{\Phi 
\tilde{d}}$, the contributions are symmetric in the interchange of $x_1$
and $x_2$, due to the scalar nature of the Higgs bosons. 

\end{itemize}

\subsection*{3.2 Integrated three--body partial widths} 

Integrating over the energies $x_1$ and $x_2$ of the two fermions, with 
boundary conditions,
\beq
1-x_1- \mu_{\chi} \leq x_2 \leq 1 - \frac{ \mu_{\chi} }{1-x_1} \ \ ,
\ \ 0 \leq x_1 \leq 1- \mu_{\chi}
\eeq
one obtains the partial decay width, which is given by an expression similar 
to eq.~(3.4): 
\beq
\Gamma_{\chi_i} = \frac{\alpha^2 N_c } {32 \pi }  m_{\chi_i} \bigg[ \Gamma_V + 
\Gamma_{\tilde{u}} + \Gamma_{\tilde{d}} + \Gamma_{\Phi}  + \Gamma_{H_1 H_2}
+ \Gamma_{V \tilde{u}} +  
\Gamma_{V \tilde{d}} + \Gamma_{\tilde{u} \tilde{d}} + \Gamma_{\Phi 
\tilde{u}} + \Gamma_{\Phi \tilde{d}}  \bigg]
\eeq
Using the phase space functions $\lambda_k$ and the function ${\cal L}_k$
defined by:
\beq
\lambda_k=1-2\mu_\chi -2\mu_k +(\mu_k-\mu_\chi)^2 \hspace*{2cm} \\
{\cal L}_k=\frac{2}{\sqrt{-\lambda_k}}
\left[ {\rm Arctan} \left(\frac{-1+\mu_\chi-\mu_k}{\sqrt{-\lambda_k}}\right) -
{\rm Arctan} \left(\frac{1-\mu_\chi-\mu_k}{\sqrt{-\lambda_k}}\right) \right]
\eeq
one has for the various contributions:
\beq
\Gamma_V &=& 8
\left[
 (v_V^f)^2 + (a_V^f)^2\right] 
\left[ (G^L_{jiV})^2 + (G^R_{jiV})^2 \right]
\Bigg\{
\frac{\mu_\chi-1} {6 \mu_V} \bigg( \lambda_V 
+ \mu_V (5+ 5\mu_\chi -7\mu_V)
\bigg) 
\non
\\
&&
- \frac{\mu_V}{2} (1 + \mu_\chi-\mu_V)
{\rm Log} \mu_\chi 
-\frac{\mu_V}{2}(\lambda_V+2\mu_{\chi}) {\cal L}_V
\Bigg\}
-8 \left[ (v^f_V)^2 + (a^f_V)^2 \right] 
G_{jiV}^L G_{jiV}^R \sqrt{ \mu_\chi}
\non
\\
&&
\Bigg\{ 
4 (\mu_\chi-1) + (1+ \mu_\chi -2\mu_V) {\rm Log} \mu_\chi
+ \bigg( 
\lambda_V -\mu_V(1+ \mu_\chi -\mu_V) 
\bigg){\cal L}_V   
\Bigg\}
\eeq
\beq
\Gamma_{\tilde{f}} &= & \sum_{k,l=1}^2 
\big( a^f_{ik} a^f_{il} + b^f_{ik} b^f_{il} \big) 
\big( a^f_{jk} a^f_{jl} + b^f_{jk} b^f_{jl} \big)
\Bigg\{ (1-\mu_\chi) (\mu_{\tilde{f}_k}  
+ \mu_{\tilde{f}_l}) -\frac{3}{2} (1- \mu_\chi^2)
\non\\
&& 
+ \frac{(\mu_{\tilde{f}_l}-1)^2(\mu_{\tilde{f}_l}-\mu_\chi)^2}{
\mu_{\tilde{f}_l}( \mu_{\tilde{f}_l} -\mu_{\tilde{f}_k})} {\rm Log} 
\frac{\mu_{\tilde{f}_l}-1}{\mu_{\tilde{f}_l} -\mu_\chi} 
+ \frac{(\mu_{\tilde{f}_k}-1)^2(\mu_{\tilde{f}_k}-\mu_\chi)^2}{
\mu_{\tilde{f}_k}( \mu_{\tilde{f}_k} -\mu_{\tilde{f}_l})} {\rm Log} 
\frac{\mu_{\tilde{f}_k}-1}{\mu_{\tilde{f}_k} -\mu_\chi}
\non\\
&& 
- \frac{\mu^2_\chi} {\mu_{\tilde{f}_k} \mu_{\tilde{f}_l}} {\rm Log}\mu_\chi
 \Bigg\}
\label{amps}
\eeq
\beq
\Gamma_{\Phi} &=& \sum_{k}2
\left[ (v_k^f)^2 + (a_k^f)^2 \right]
\left[(G_{ijk}^L)^2+(G_{ijk}^R)^2 \right]
 \Bigg\{\frac{1}{2}(1-\mu_\chi) (6\mu_k -5 -5\mu_\chi)
\non\\
&&  +\frac{1}{2} \left[ -5\mu_\chi^2 \mu_k - 3\mu_k^3 + 7\mu_k^2 +1
-\mu_\chi^2-\mu_\chi +\mu_\chi^3 -5\mu_k + 7\mu_\chi \mu_k^2 
- 2\mu_\chi \mu_k \right] {\cal L}_k 
\non\\
&& + \frac{1}{2}\left(1-4\mu_k-4\mu_\chi \mu_k + 3\mu_k^2 
+\mu_\chi^2\right) {\rm Log}\mu_\chi  \Bigg\}
+ 4 \left[ (v_k^f)^2 + (a_k^f)^2 \right]  G_{ijk}^L 
G_{ijk}^R \sqrt{ \mu_\chi} 
\non\\
&& \Bigg\{ 4 (\mu_\chi-1) + (1+ \mu_\chi -2\mu_k) {\rm Log} \mu_\chi
+ \bigg( \lambda_k -\mu_k(1 + \mu_\chi - \mu_k) \bigg)
{\cal L}_k
\Bigg\}
\eeq
\beq
\Gamma_{H_1 H_2} &=& 2 v_1^f v_2^f \Bigg\{ 
\bigg( G^L_{ij1} G^L_{ij2} + G^R_{ij1} G^R_{ij2} \bigg) 
\bigg[ (2\mu_{H_1}+2\mu_{H_2}-3\mu_\chi-3)(1-\mu_\chi) 
\non\\
&&                                                                             
+ \frac{\mu_{H_1}(1+\mu_\chi-\mu_{H_1})}{\mu_{H_2}-\mu_{H_1}} 
\lambda_{H_1} {\cal L}_{H_1} 
- \frac{\mu_{H_2}(1+\mu_\chi-\mu_{H_2})}
{\mu_{H_2}-\mu_{H_1}} \lambda_{H_2} {\cal L}_{H_2}
\non\\
&&
+ \bigg(1 +\mu_\chi^2 +\mu_{H_1}^2 
+\mu_{H_2}^2 +\mu_{H_1}\mu_{H_2} -2(1+\mu_\chi)(\mu_{H_1}+\mu_{H_2})\bigg) 
{\rm Log}\mu_\chi  \bigg] 
\non\\
&& + 2\sqrt{\mu_\chi} \bigg( G^L_{ij1} G^R_{ij2} +G^L_{ij2}  G^R_{ij1} 
\bigg) 
\bigg[ \frac{\mu_{H_1}}{\mu_{H_2}-\mu_{H_1}}
\lambda_{H_1} {\cal L}_{H_1} - \frac{\mu_{H_2}}{\mu_{H_2}-\mu_{H_1}}
\lambda_{H_2} {\cal L}_{H_2} 
\non\\
&& + \frac{\mu_{H_1}^2-\mu_{H_2}^2-(\mu_{H_1}
-\mu_{H_2})(1+\mu_\chi)}{\mu_{H_2}-\mu_{H_1}} {\rm Log} \mu_\chi 
-2(1-\mu_\chi) \bigg] \Bigg\}
\eeq
\beq
\Gamma_{V\tilde f} &=& 4\sum_{k=1}^2 \Bigg\{
A_1^f \Bigg( 
\frac{\mu_{\chi}-1}{4}(\mu_{\chi}+1-4\mu_{\tilde f_k}+2\mu_V)
+\frac{1}{4} (1+\mu_{\chi}-2\mu_{\tilde f_k}+\mu_V) 
\lambda_{V} {\cal L}_{V}
\non\\
&& +\frac{1}{4} (1+4\mu_{\chi}+\mu_{\chi}^2-2\mu_{\tilde f_k}
-2\mu_{\tilde f_k}\mu_\chi+2\mu_{\tilde f_k} \mu_V-\mu_V^2) {\rm Log} 
\mu_{\chi} +(\mu_\chi-\mu_{\tilde{f}_k})(-1+\mu_{\tilde{f}_k})
\non \\
&&  {\cal F}(a^V_+,a^V_-, \mu_{\tilde f_k},\mu_V) \Bigg) 
-A_2^f \sqrt{\mu_\chi}\Bigg(
\mu_{\chi}-1 -\frac{\mu_{\chi}}{\mu_{\tilde f_k}}{\rm Log} \mu_{\chi}
+ \mu_V {\cal F} (a^V_+,a^V_-, \mu_{\tilde f_k},\mu_V)
\non \\
&& -\frac{1}{\mu_{\tilde f_k}}
(\mu_{\chi}-\mu_{\tilde f_k}-\mu_{\chi}\mu_{\tilde f_k}+\mu_{\tilde f_k}^2)
{\rm Log} \frac{\mu_{\tilde f_k}-1}{\mu_{\tilde f_k}-\mu_{\chi}}
\Bigg)\Bigg\}
\eeq
where
\beq
A_1^d &=& - [ a^d_{ik} a^d_{jk} G_{jiV}^R(v^d_V+a^d_V)
+b^d_{ik} b^d_{jk} G_{jiV}^L(v^d_V-a^d_V) ] \non\\
A_1^u &=&  a^u_{ik} a^u_{jk} G_{jiV}^L(v^u_V+a^u_V)
+b^u_{ik} b^u_{jk} G_{jiV}^R(v^u_V-a^u_V)  \non\\
A_2^d &=& - [ a^d_{ik} a^d_{jk} G_{jiV}^L (v^d_V+a^d_V)  
+b^d_{ik} b^d_{jk} G_{jiV}^R (v^d_V-a^d_V) ] \non\\
A_2^u &=& a^u_{ik} a^u_{jk} G_{jiV}^R (v^u_V+a^u_V)  
+b^u_{ik} b^u_{jk} G_{jiV}^L (v^u_V-a^u_V)
\eeq
\beq
\Gamma_{\Phi \tilde f} &=& \sum_{k,l} \Bigg\{
B_1^f
\Bigg(
(1-\mu_{\chi})(-1+2\mu_k-\mu_{\chi}+2\mu_{\tilde f_l})
-\mu_k \lambda_k {\cal L}_k -\mu_k (1+\mu_\chi-\mu_k){\rm Log} \mu_{\chi} \non 
\\
&&
-2{\rm Log}\frac{\mu_{\tilde f_l}-\mu_{\chi}}{\mu_{\tilde f_l}-1}
(\mu_{\chi}-\mu_{\tilde f_l}-\mu_{\chi}\mu_{\tilde f_l}+\mu_{\tilde f_l}^2)
-2\mu_k \mu_{\tilde f_l} {\cal F}( a_+^{H_k}, a_-^{H_k}, \mu_{\tilde f_l},\mu_k) \Bigg)
\non
\\
&&
-2 B_2^f \sqrt{\mu_{\chi}} 
\Bigg(\mu_{\chi}-1-\frac{\mu_{\chi}}{\mu_{\tilde f_l}}{\rm Log} \mu_{\chi}
-\frac{1}{\mu_{\tilde f_l}} ( \mu_{\chi} -\mu_{\tilde f_l} -
\mu_{\chi}\mu_{\tilde f_l} +\mu_{\tilde f_l}^2 )
{\rm Log}\frac{\mu_{\tilde f_l}-1}{\mu_{\tilde f_l}-\mu_{\chi}}
\non \\
&& + \mu_k {\cal F}( a_+^{H_k}, a_-^{H_k}, \mu_{\tilde f_l},\mu_k) 
\Bigg) \Bigg\}
\eeq
where
\beq
B_1^d &=& (v^d_k-a^d_k)a^d_{il}b^d_{jl}G^R_{ijk}
+(v^d_k+a^d_k)b^d_{il}a^d_{jl}G^L_{ijk}  \non\\
B_1^u &=& (v^u_k-a^u_k)a^u_{jl}b^u_{il}G^R_{ijk}
+(v^u_k+a^u_k)a^u_{il}b^u_{jl}G^L_{ijk} \non\\
B_2^d &=& (v^d_k-a^d_k)a^d_{il}b^d_{jl}G^L_{ijk}
+(v^d_k+a^d_k)b^d_{il}a^d_{jl}G^R_{ijk} \non\\
B^2_u &=& (v^u_k-a^u_k) b^u_{il} a^u_{jl} G^L_{ijk} + 
(v^u_k+a^u_k) a^u_{il} b^u_{jl} G^R_{ijk}
\eeq
\beq
\Gamma_{\tilde u\tilde d} &=&-2\sum_{k,l=1}^{2}\Bigg\{  
( a^u_{jk}a^d_{il} b^u_{ik} b^d_{jl} + a^u_{ik} a^d_{jl} b^u_{jk} b^d_{il} ) 
\Bigg(
\frac{1}{2} (\mu_{\chi}-1)(2\mu_{\tilde u_k}+2\mu_{\tilde d_l}-\mu_{\chi}-1)
-\mu_{\chi}{\rm Log}\mu_{\chi}
\non\\
&& +{\rm Log} \frac{\mu_{\tilde u_k}-1}{\mu_{\tilde u_k}-\mu_{\chi}}
(\mu_{\tilde u_k}-\mu_{\chi})(1-\mu_{\tilde u_k})
+ {\rm Log}\frac{\mu_{\tilde d_l}-1}{\mu_{\tilde d_l}-\mu_{\chi}}
(\mu_{\tilde d_l}-\mu_{\chi})(1-\mu_{\tilde d_l}) 
\non \\
&& + (\mu_\chi-\mu_{\tilde u_k}\mu_{\tilde d_l})
\tilde{{\cal F}}((\mu_{\tilde u_k}-\mu_\chi)/\mu_{\tilde u_k},\mu_{\tilde u_k}-\mu_\chi,\mu_{\tilde d_l},\mu_{\tilde u_k}) \Bigg) 
+ 
( a^u_{ik} a^u_{jk} a^d_{il} a^d_{jl} + b^u_{ik} b^u_{jk} b^d_{il} b^d_{jl} )
\non\\
&& 
\sqrt{\mu_\chi}  \Bigg(
{\rm Log} \frac{\mu_{\tilde u_k}-1}{\mu_{\tilde u_k}-\mu_{\chi}}
(\mu_{\tilde u_k}-\mu_{\chi})(1-\mu_{\tilde u_k})/\mu_{\tilde u_k} +
{\rm Log}\frac{\mu_{\tilde d_l}-1}{\mu_{\tilde d_l}-\mu_{\chi}}
(\mu_{\tilde d_l}-\mu_{\chi})(1-\mu_{\tilde d_l})/\mu_{\tilde d_l} 
\non\\
&&
+2(\mu_{\chi}-1)
+ (1+\mu_\chi-\mu_{\tilde d_l}-\mu_{\tilde u_k})
\tilde{{\cal F}}((\mu_{\tilde u_k}-\mu_\chi)/\mu_{\tilde u_k},\mu_{\tilde u_k}-\mu_\chi,\mu_{\tilde d_l},\mu_{\tilde u_k})
\non\\
&&  
-\bigg(\frac{\mu_{\chi}}{\mu_{\tilde u_k}}+\frac{\mu_{\chi}}{\mu_{\tilde d_l}}
\bigg){\rm Log}\mu_{\chi}
 \Bigg) \Bigg\}
\label{ampi}
\eeq 
In the previous expressions, we have used the variables and functions:   
\beq
a^i_\pm =\frac{1}{2} (1-\mu_{\chi}+\mu_i \pm \sqrt{\lambda_i})
\eeq
\beq
{\cal F}(a,b, \mu_i,\mu_j)= f(a, \mu_i)+f(b, \mu_i)-f(1,\mu_i)+
{\rm Log}\mu_j {\rm Log}\frac{\mu_i-\mu_\chi}{\mu_i-1} \non\\
\tilde{{\cal F}}(a,b, \mu_i,\mu_j)= f(a, \mu_i)-f(b, \mu_i)-f(1,\mu_i)-
{\rm Log}\mu_j {\rm Log}\frac{\mu_i-\mu_\chi}{\mu_i-1}
\eeq
\beq
f(a, \mu_i)={\rm Li}_2 \left(\frac{\mu_i-\mu_\chi}{a+\mu_i-1}\right)
-{\rm Li}_2 \left( \frac{\mu_{i}-1}{a+\mu_i-1} \right) -
{\rm Log}(a+\mu_i-1){\rm Log}\left(\frac{\mu_i-\mu_{\chi}}{\mu_i-1}  \right)
\eeq
where ${\rm Li}_{2}$ is the Spence function defined by ${\rm Li}_2(x) = 
\int_0^1 t^{-1} {\rm Log}(1-xt)dt$. 

\subsection*{3.3 The two--body partial decay widths}

The two--body partial decay widths can be obtained from the expressions given
in section 3.1 by including the total decay widths of the exchanged gauge
and Higgs bosons and the sfermions. In this case a smooth transition between
three-- and two--body partial decay widths can be obtained. We will list
below the integrated form of the two--body partial decay widths of charginos
and neutralinos into sfermion--fermion pairs [with massive fermions], and into 
neutralino and gauge or Higgs boson final states; see also Ref.~\cite{two-body}.
\beq
\Gamma(\chi_i \ra f  \tilde{f}_j) = \frac{\alpha N_c}{8}\, m_{\chi_i} \, \bigg[
 \left( (a_{ij}^f)^2 + (b_{ij}^f)^2 \right) (1- \mu_{\tilde f_j} + \mu_f)
+ 4 \sqrt{\mu_f} a_{ij}^f b_{ij}^f \bigg] \, \lambda^{\frac{1}{2}} (\mu_f, 
\mu_{\tilde f_j})
\eeq
\beq
\Gamma(\chi_i \ra \chi_j V) = \frac{\alpha}{8} \, m_{\chi_i} \, 
\lambda^{\frac{1}{2}}(\mu_{\chi_j},\mu_V) \left\{ -12 \sqrt{\mu_{\chi_j}}
G_{jiV}^L G_{jiV}^R \right. \hspace*{3.8cm} \non \\
\hspace*{.9cm} + \left. \left[ (G_{jiV}^L)^2 + (G_{jiV}^R)^2 \right] 
(1+ \mu_{\chi_j}-\mu_V) +(1- \mu_{\chi_j} +\mu_V)(1- \mu_{\chi_j}-\mu_V) 
\mu_V^{-1} \right\}
\eeq
\beq
\Gamma(\chi_i \ra \chi_j H_k) = \frac{\alpha}{8} \, m_{\chi_i} \, 
\lambda^{\frac{1}{2}}(\mu_{\chi_j},\mu_{H_k}) \left\{ \left[ (G^L_{ijk})^2
+(G^R_{ijk})^2 \right] ( 1+ \mu_{\chi_j} -\mu_{H_k}) \right. \non \\
+\left. 4 \sqrt{\mu_{\chi_j}} \, G^L_{ijk} G^R_{ijk}  \right\} \hspace*{7cm}
\eeq
with 
\beq
\lambda (x,y)=1+ x^2+y^2-2x-2y-2xy \ \ \ \ , \ \ \ \mu_X= m_X^2/m_{\chi_i}^2
\eeq

\subsection*{3.4 Decays into gluino and quark--antiquark final states}

As discussed in section 2.1, in models without gaugino mass unification at the 
GUT scale, the lightest chargino and the next--to--lightest neutralino could
be heavier than the gluino. In this case, the three--body decay modes
\beq
\chi_i \ \to \ \tilde{g} \ u \, \bar{d} 
\eeq
with $\chi_i \equiv \chi_1^\pm$ or $\chi_2^0$, are kinematically accessible. This
decay is mediated by $t$-- and $u$--channel exchange of squarks only; Fig.~5. 

\vspace*{-1cm}
\begin{picture}(1000,200)(10,0)
\Text(140,130)[]{$\chi_i$}
\ArrowLine(110,120)(160,120)
\ArrowLine(160,120)(190,150)
\Text(197,150)[]{$\bar{d}$}
\DashArrowLine(160,120)(180,90){4}
\Text(155,100)[]{$\tilde{q}$}
\ArrowLine(180,90)(210,120)
\Text(215,120)[]{$u$}
\ArrowLine(200,65)(180,90)
\Text(210,65)[]{$\tilde{g}$}
\Text(300,130)[]{$\chi_i$}
\ArrowLine(260,120)(310,120)
\ArrowLine(310,120)(340,150)
\Text(347,150)[]{$u$}
\DashArrowLine(310,120)(330,90){4}{}
\Text(305,100)[]{$\tilde{q}$}
\ArrowLine(330,90)(360,120)
\Text(370,120)[]{$\bar{d}$}
\ArrowLine(350,65)(330,90)
\Text(360,65)[]{$\tilde{g}$}
\end{picture}
\vspace*{-2.5cm}

\begin{center}
\nn {\it Figure 5: The Feynman diagrams contributing to the three--body decay 
$\chi_i \to \tilde{g} q \bar{q}$.} 
\end{center}
\vspace*{4mm} 
\setcounter{figure}{5}

The Dalitz density and the partial decay width, neglecting the masses of the 
final state quarks, are given by: 
\beq
\frac{ \dx \Gamma_{\chi_i} } {\dx x_1 \dx x_2}  &=& \frac{e^2 g_s^2 m_{\chi_i}} 
{8 (2\pi)^3}  \,  \bigg[ \dx \Gamma_{\tilde{u}} + \dx 
\Gamma_{\tilde{d}} + \dx \Gamma_{\tilde{u} \tilde{d}} \bigg] \non \\
\Gamma_{\chi_i} &=& \frac{\alpha \alpha_s} {4 \pi }  m_{\chi_i} 
\bigg[ \Gamma_{\tilde{u}} + \Gamma_{\tilde{d}}+\Gamma_{\tilde{u} \tilde{d}} 
\bigg]
\eeq
with $x_1=2E_u/m_{\chi_i}, x_2= 2E_{d}/m_{\chi_i}$. The various amplitudes 
are as in eqs.~(\ref{amp1dd},\ref{amp1uu},\ref{amp1ud}) for the Dalitz 
densities and eqs.~(\ref{amps},\ref{ampi}) for the integrated width, with now, 
$\mu_\chi \equiv m^2_{\tilde g}/ m_{\chi_i}^2$. One has also to replace the 
final neutralino--$f$--$\tilde{f}_{l}$ couplings, $a_{jl}^f, b_{jl}^f$, by the
gluino--quark--squark couplings, $a_{l}^q, b_{l}^q$, which in the case of 
mixing read: 
\beq
a_1^q = b_2^q = \sin \theta_q \ \ , \ \ 
a_2^q = - b_1^q = \cos \theta_q 
\eeq
The expressions for the three body decays of gluinos into 
$\chi_i+ q\bar{q}$ final states \cite{gluino,N1}, are given by the previous 
formulae with the interchange of $m_{\chi_i}$ and $m_{\tilde g}$ and by 
dividing the result by a factor of $8$ to account for the color numbers of the 
gluino. [Note that this factor is missing in the expression of the gluino decay
width in Ref.~\cite{N1}; since it is a global factor, the branching ratios
are therefore not affected.]. 
\newpage

\section*{4. Numerical Analysis} 

We will first illustrate our results in an mSUGRA type model, where we assume a
universal mass $m_0$ for the scalar fermions and a mass $m_{1/2}$ for the
gauginos at the GUT scale; the soft SUSY breaking masses for the Higgs bosons
are however disconnected from the one of the sfermions so that the
pseudo--scalar Higgs boson mass $M_A$ and the higgsino parameter $\mu$ are free
parameters [in contrast to the mSUGRA model where $\mu$ is determined, up to
its sign, from the requirement of electroweak symmetry breaking]. For the
squark sector, we will use the simple expressions eqs.~(\ref{smass}) for the
soft SUSY breaking left-- and right--handed squark and slepton masses when
performing the RGE evolution to the weak scale at one--loop order if the Yukawa
couplings in the RGE's are neglected\footnote{As mentioned previously, for
third generation sfermions, neglecting the Yukawa couplings in the RGE is  a
poor approximation since these couplings can be large; this is particularly the
case for top squarks which however will not be considered in the present
analysis, since we will assume that charginos and neutralinos are not heavy
enough to decay into top quark final states.}. One has then, in the case of the
third generation sparticles, to include the mixing. Since for sbottoms and
stau's, large enough off--diagonal elements of the mass matrices are obtained
only for large $\mu$ and $\tb$ values and the trilinear couplings play only a
marginal role, we will fix the latter to $A_b=A_\tau=-500$ GeV in the entire
analysis. We will choose two representative values for $\tb$: a ``low" value
($\tb=5$) and a large value ($\tb=50)$ and two values for the pseudoscalar $A$
boson mass\footnote{In the large $\tb$ scenario and in the non--decoupling 
regime, the experimental bounds on the masses of the pseudoscalar Higgs boson 
$A$ and the lightest Higgs boson $h$ in the MSSM from negative searches at 
LEP2 are $M_A, M_h \gsim 93.5$ GeV \cite{R7}.}, $M_A =100$ and 500 GeV. \s
 
In a second step, we will relax the gaugino mass unification constraint
$m_1=m_2=m_3=m_{1/2}$ at the GUT scale, and use the weak scale gaugino masses
$M_1$ and $M_2$ given in Table 1 for the $F_\Phi$ representations and the {\bf
OII} model. We will still use the soft--SUSY breaking scalar masses given in
eq.~(\ref{smass}). In this case, we will stick in the illustrations to
the large $\tb$ scenario, $\tb=50$, but still show the effect of the Higgs 
boson contribution by taking the two examples  $M_A=100$ and 500 GeV. \s

In most of the cases, the wino mass parameter will be fixed to $M_2=150$ GeV, 
which for large values of $\mu$, leads in an mSUGRA--type model to the masses 
$m_{\chi_1^\pm} \simeq m_{\chi_2^0} \simeq  150$ GeV and $m_{\chi_1^0} \simeq 
75$ GeV [there is a very small variation with the value of $\tb$] and hence to 
states which are accessible at the high--luminosity phase of the Tevatron and 
at a future $e^+e^-$ linear collider with a c.m. energy of 500 GeV. \s 

Note that in the entire analysis, we will include the radiative corrections to
the $b$--quark and $\tau$--lepton masses, as well as the radiative corrections
to the chargino, neutralino and gluino masses given in section 2.1.  We will
also take into account the full dependence on the final state fermion masses
[using the pole masses $m_b=4.6$ GeV, $m_\tau=1.78$ GeV and $m_c=1.45$ GeV in
the phase space, the other fermions are taken to be massless] since in some
cases [in particular when the decaying chargino or neutralino has a mass which
is close to the final LSP mass], they play a significant role. \s

The branching ratios for the lightest chargino $\chi_1^\pm$ and 
next--to--lightest neutralino $\chi_2^0$ into the LSP and $\tau$ and 
$b$--quark final states are shown in Figs.~6--8, in model {\bf 1} with
gaugino mass unification at $M_{\rm GUT}$.  The wino mass parameter is
fixed to $M_2=150$ GeV and the choices $\tb=5,50$ and $M_A=100,500$ GeV 
have been made. \s

In Fig.~6a, BR$(\chi_1^+ \to \chi_1^0 \tau^+ \nu)$ is shown as a function of
the lightest $\tilde{\tau}_1$ mass for $\mu=+500$ GeV. For large values of
$m_{\tilde{\tau}_1}$ and with a heavy charged Higgs boson $[M_A=500$ GeV
leading to $M_{H^\pm}=506$ GeV], the branching ratio is small, being at the
level of $10\%$. In this regime, the dominant contribution is coming from the
virtual $W$ boson exchange and  BR$(\chi_1^+)$ is practically the same as BR($W
\to f\bar{f})$, i.e. $\sim 10\%$ for the $\tau^+ \nu$ final state.  However,
for large values of $\tb$ and for a light $H^\pm$ boson $[M_A=100$ GeV leading
to $M_{H^\pm}\sim 128$ GeV], the charged Higgs boson contribution [since the
$H^\pm \nu \tau^\mp$ couplings are enhanced] becomes dominant and the fraction
BR$(\chi_1^+ \to \chi_1^0 \tau^+ \nu)$ can reach the level of 40\% even for
$m_{\tilde{\tau}_1} \sim 500$ GeV. For smaller values of $m_{\tilde{\tau}_1}$,
the virtual stau exchange diagram becomes more and more dominant, and 
BR$(\chi_1^+ \to \chi_1^0 \tau^+ \nu)$ becomes close to $\sim 80\%$ for stau
masses of the order of 150 GeV. If in addition, $H^\pm$ is relatively light,
the branching ratio reaches the level of 100\%.  \s

Figs.~6b and 6c, where BR$(\chi_1^+ \to \chi_1^0 \tau^+ \nu)$ is plotted 
for a common sfermion mass $m_0=300$ GeV as a function of $\mu$ and $\tb$, 
respectively, show the same trend from a different perspective. For small 
values of $\tb$, the mixing in the stau sector and the Yukawa couplings of the 
$\tau$ lepton are not enhanced and the branching fraction is at the level of 
$10\%$. But for large $\tb$ values, the stau becomes light and the branching 
ratio becomes close to unity for large values of $\mu$. This occurs more
quickly, if the charged Higgs boson is light. \s

Figs.~7 and 8 show, respectively, the branching ratios BR$(\chi_2^0 \to
\chi_1^0 \tau^+ \tau^-)$ and BR$(\chi_2^0 \to \chi_1^0 b \bar{b})$, as
functions of $m_{\tilde{\tau}_1}$ or $m_{\tilde{b}_1}$ for $\mu=500$ GeV (a),
as a function of $\mu$ (b) and as a function of $\tb$ (c) for $m_0=300$ GeV. 
In this case, there is a competition between $b\bar{b}$ and $\tau^+ \tau^-$
final states. In the case of a light $A$ boson and for large $\tb$ values, the
$A$ and $h$ contributions are much more important in the decay $\chi_2^0 \to
\chi_1^0 b \bar{b} $ than in the channel $\chi_2^0 \to \chi_1^0 \tau^+ \tau^-$
because of the larger $b$--quark mass and the color factor; the Higgs
contribution makes then BR$(\chi_2^0 \to \chi_1^0 b \bar{b})$ dominating,
except when $\tilde{\tau}_1$ is very light, and the two body decay $\chi_2^0
\to \tilde{\tau}_1 \tau$ is close to occur, making  BR$(\chi_2^0 \to \chi_1^0
\tau^+ \tau^-)$ close to unity. Even for  heavy $A,H$ bosons, BR$(\chi_2^0 \to
\chi_1^0 b \bar{b})$ can reach the level of $\sim 50\%$.  However, for large
enough values of $\tb$ and $\mu$, it is the decay channel $\chi_2^0 \to
\chi_1^0 \tau^+ \tau^-$ which dominates, since for a universal scalar mass
$m_0$, the stau is always lighter than the $\tilde{b}_1$ state and its virtual
contribution is larger, despite of the color factor. Needless to say, the sum
of the two branching ratios, BR$(\chi_2^0 \to \chi_1^0 \tau^+ \tau^- + \chi_1^0
b \bar{b})$ is in general close to unity. \s 

In Fig.~9, we illustrate the effect of the radiative corrections to the
$b$--quark mass [and to a lesser extent the tau--lepton mass] by showing the
branching ratios BR$(\chi_2^0 \to \chi_1^0 \tau^+ \tau^-)$ and BR$(\chi_2^0 \to
\chi_1^0 b \bar{b})$ as a function of $\tb$ with $\mu, m_0, M_A$ and $M_2$
fixed to, respectively, the values 1 TeV, 300 GeV, 150 and 150 GeV.
For $\mu >0 \, (<0)$, the SUSY radiative corrections [in particular, the
correction due to sbottom--gluino loops] decrease (increase) substantially the
value of $m_b$, therefore suppressing (enhancing) the $\chi_2^0 \to \chi_1^0 b
\bar{b}$ rate by a sizeable factor, compared to the branching ratio without the
correction (solid lines), for large enough $\tb$ values. The fraction
BR$(\chi_2^0 \to \chi_1^0 \tau^+ \tau^-)$ increases (decreases) then,
accordingly. These corrections are therefore very important and must be taken
into account. \s

In Figs.~10, 11 and 12, we show, respectively, the branching fractions
BR$(\chi_1^+ \to \chi_1^0 \tau^+ \nu_\tau)$, BR$(\chi_2^0 \to \chi_1^0 \tau^+
\tau^-)$ and BR$(\chi_2^0 \to \chi_1^0 b \bar{b})$ as functions of $\mu (>0)$
in the models ${\bf 24, 75, 200}$ and ${\bf OII}$ without gaugino mass
unification as well as in the universal model {\bf 1} for comparison.  The
various parameters are fixed to the following values: $\tb=50$, $M_2=150$ GeV,
$m_0=500$ and $M_A=100$ (a) and 500 GeV (b).  Before discussing the various 
decay channels in these models, compared to the universal case, let us make two 
general comments: \s

$i)$ The values of $M_{1,2,3}$ at the weak scale are different and modify 
appreciably the phase space for the decays; in particular two--body decay 
modes and decays into gluinos  become possible. In addition the radiative
corrections to the gaugino masses, although only of the order of a few GeV, 
could allow the opening of channels such as those involving tau leptons. \s

$ii)$ Due to the different values of $M_{1,2}$, the evolution of the sfermion 
masses from $\Lambda_{\rm GUT}$ to the weak scale are modified, and the 
contributions of $\tau$ sleptons and bottom squarks can be enhanced or 
suppressed compared to the universal case. Also, the radiative corrections to 
the fermion masses are different and can lead to a further enhancement or 
suppression of the Higgs boson and/or sfermion contribution to the decays.

\begin{figure}[htbp]
\vspace*{-4.5cm}
\hspace*{-2.3cm}
\mbox{\psfig{figure=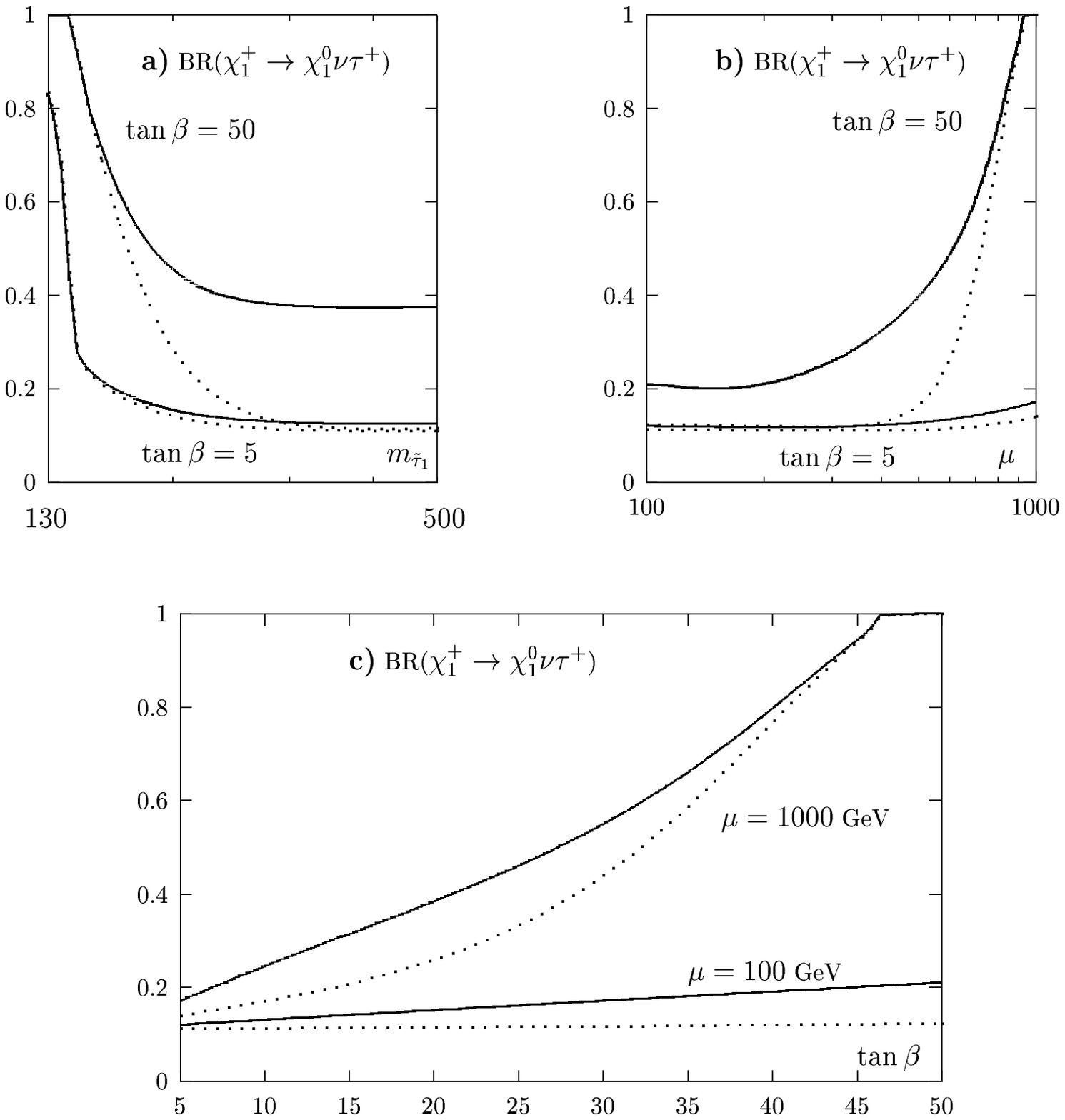,width=22.5cm}}
\vspace*{-10.cm}
\caption[]{\it The branching ratio BR$(\chi^+_1 \ra \chi_1^0 \nu \tau^+)$ for
two values of $\tb=5$ and 50 and two values of $M_A=100$ GeV (solid lines) and 
$500$ GeV (dashed lines) as a function of $m_{\tilde{\tau}_1}$ for $\mu=500$ 
GeV (a) as a function of $\mu$ assuming $m_{0}=300$ GeV (b) and as a function 
of $\tb$ for two values of $\mu=100$ and $1000$ GeV (c); $M_2$ is fixed to 
$150$ GeV.}  
\end{figure}

\begin{figure}[htbp]
\vspace*{-4.5cm}
\hspace*{-2.3cm}
\mbox{\psfig{figure=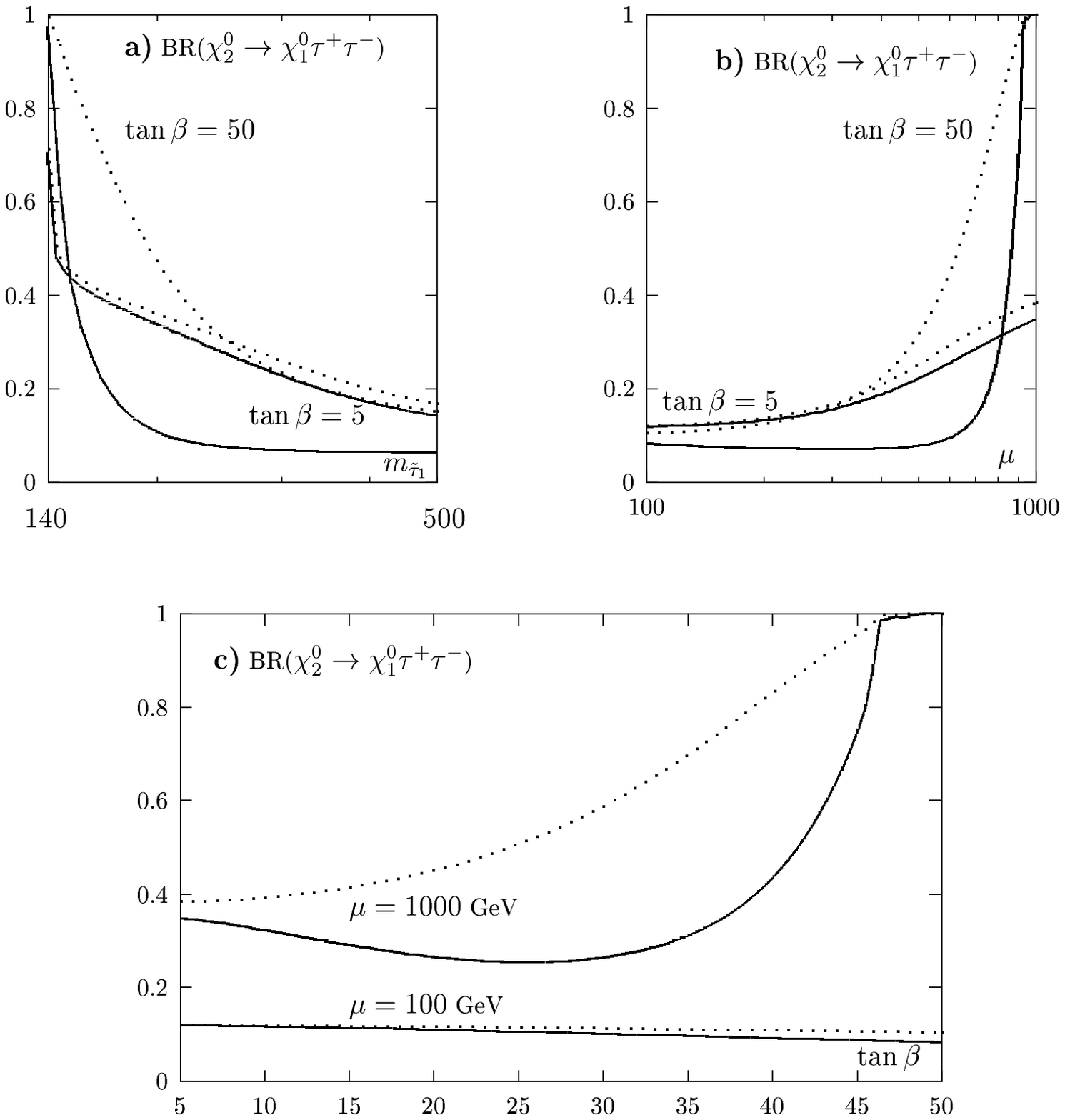,width=22.5cm}}
\vspace*{-10.cm}
\caption[]{\it The branching ratio BR$(\chi^0_2 \ra \chi_1^0 \tau^+ \tau^-)$ 
for two values of $\tb=5$ and 50 and two values of $M_A= 100$ GeV (dashed 
lines) and $500$ GeV (solid lines) as a function of $m_{\tilde{\tau}_1}$ for 
$\mu=500$ GeV (a) as a function of $\mu$ assuming  $m_{0}=300$ GeV (b) and as 
a function of $\tb$ for two values of $\mu=100$ and $1000$ GeV (c); $M_2$ is 
fixed to $150$ GeV.}  
\end{figure}

\begin{figure}[htbp]
\vspace*{-4.5cm}
\hspace*{-2.3cm}
\mbox{\psfig{figure=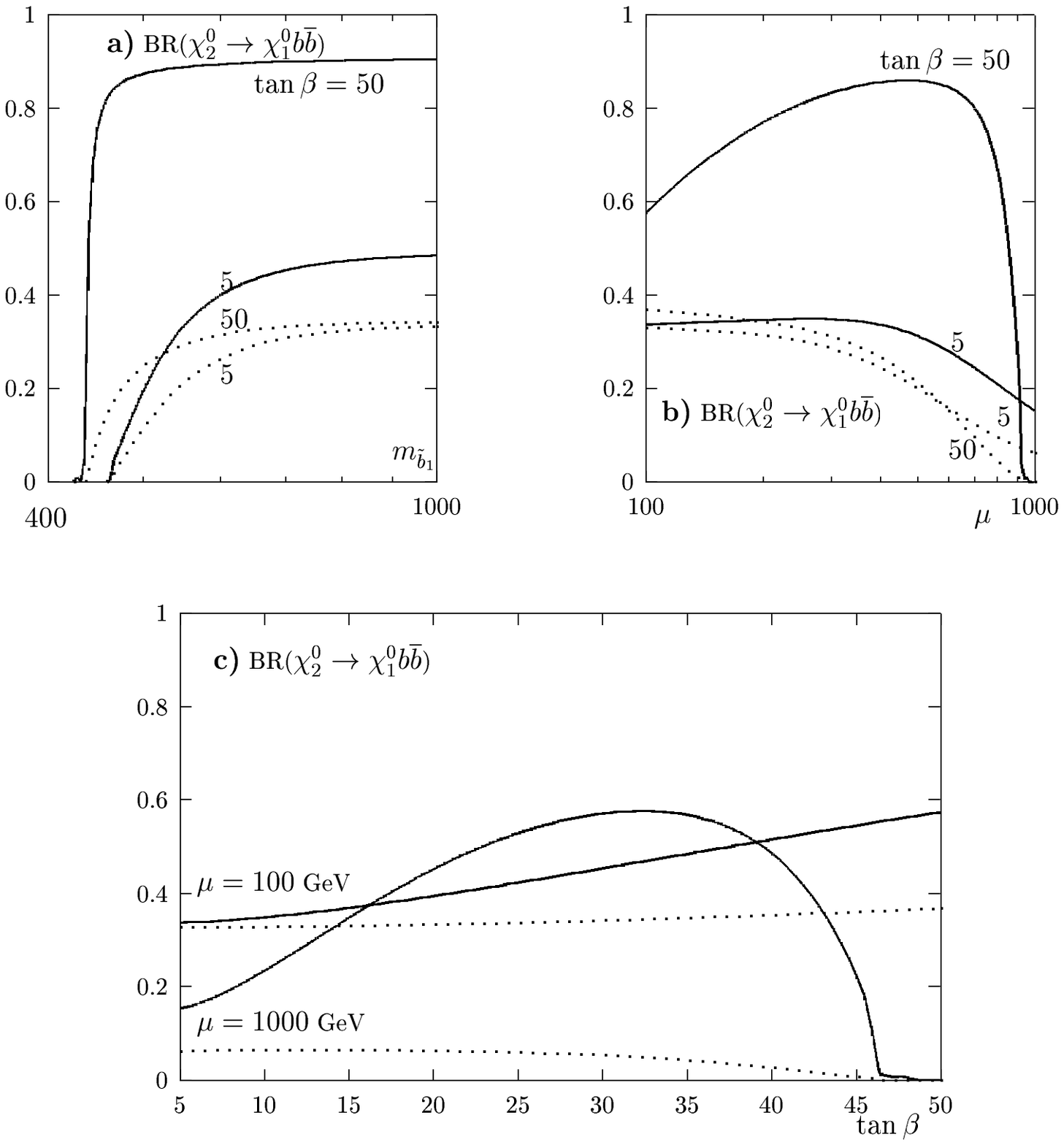,width=22.5cm}}
\vspace*{-10.cm}
\caption[]{\it The branching ratio BR$(\chi^0_2 \ra \chi_1^0 \bar{b}b)$ for
two values of $\tb=5$ and 50 and two values of $M_A= 100$ GeV (solid lines) 
and $500$ GeV (dashed lines) as a function of $m_{\tilde{b}_1}$ for $\mu=500$ 
GeV (a) as a function of $\mu$ assuming  $m_{0}= 300$ GeV (b) and as a function
of $\tb$ for two values of $\mu=$ 100 and $1000$ GeV (c); $M_2$ is fixed to 
$150$ GeV.}  
\end{figure}

\begin{figure}[htbp]
\vspace*{-4.5cm}
\hspace*{-2.3cm}
\mbox{\psfig{figure=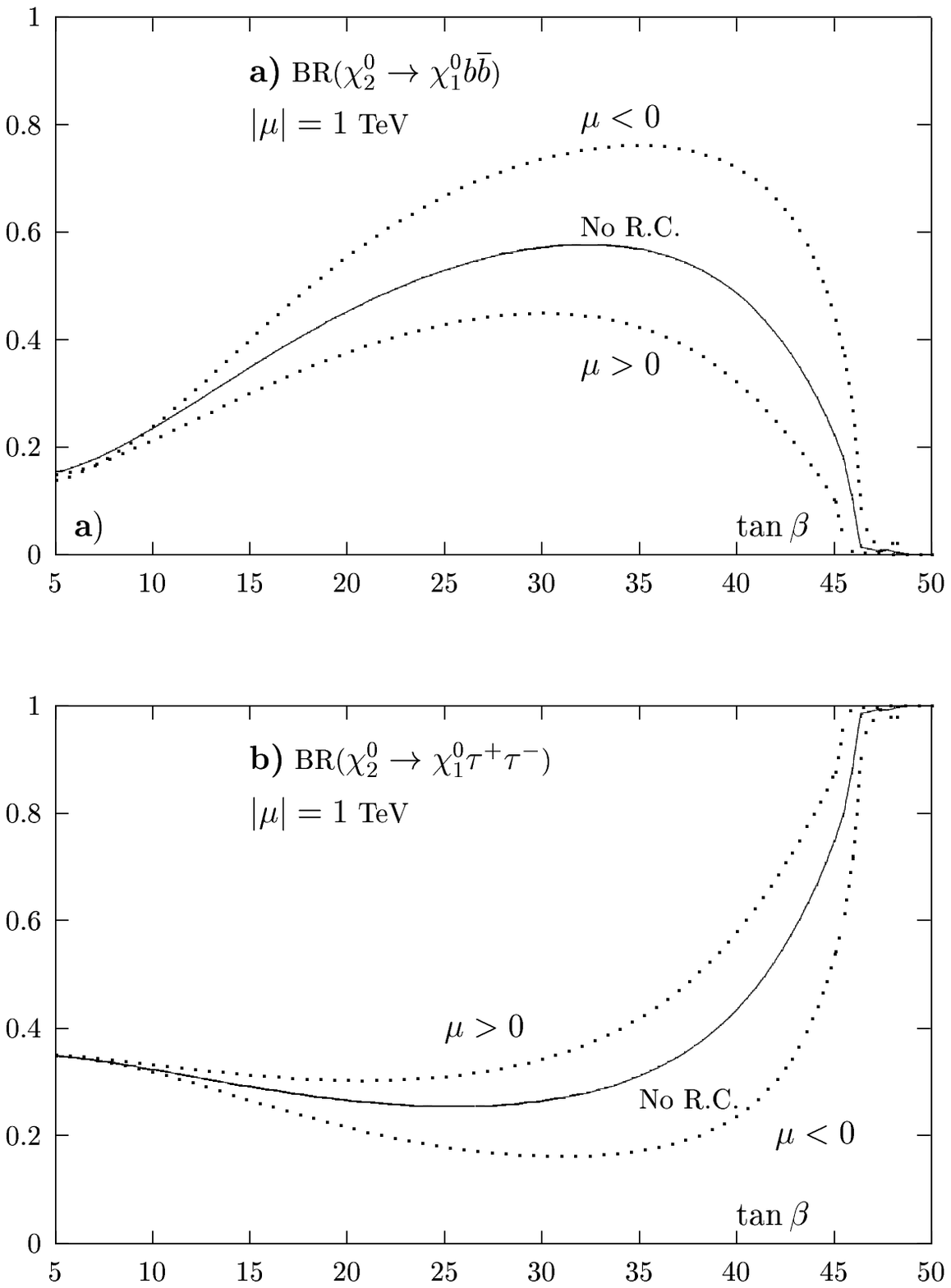,width=22.5cm}}
\vspace*{-10.cm}
\caption[]{\it The branching ratios BR$(\chi^0_2 \ra \chi_1^0 b \bar{b}$ (a)
and BR$(\chi^0_2 \ra \chi_1^0 \tau^+ \tau^-)$ (b) as a function of $\tb$ for
$|\mu|=1$ TeV, $M_A=150$ GeV, $m_0=300$ GeV and  $M_2=150$ GeV, with and 
without the radiative corrections to the fermion masses.} 
\end{figure}

{\bf Model 24}: For large $\mu$ values, $\mu \gsim 200$ GeV, the lightest 
chargino and neutralinos are gaugino like and because $M_2 \sim 6 M_1$, 
the mass differences $m_{\chi_1^+} - m_{\chi_1^0}$ and $m_{\chi_2^0} - 
m_{\chi_1^0}$ are large, making the decays into real gauge bosons, $\chi_1^+ 
\to \chi_1^0 W$ and $\chi_2^0 \to \chi_1^0 Z$, kinematically possible. The
branching ratios for $\chi_1^+$ and $\chi_2^0$ are then controlled by the
$W/Z$ branching ratios: BR$(W \to \tau^+ \nu) \sim 10\%$, BR$(Z \to \tau^+ 
\tau^-) \sim 3\%$ and BR$(Z \to b\bar{b}) \sim 15\%$. For smaller $\mu$ values,
$\mu \lsim 200$ GeV, the two neutralinos are mixtures of gauginos and higgsinos
and three--body decays are possible.  The sfermion exchange channels increase
the rates for the $\chi_2^0 \to \chi_1^0 \tau^+ \tau^-$ and $\chi_1^0 b\bar{b}$
decay channels, with an additional enhancement, in the later channel, being to
the exchange of the light Higgs bosons for $M_A, M_h \sim 100$ GeV [this 
contribution is milder in the case of $\tau^+ \tau^-$ final states because of 
the reduced Yukawa coupling]. \s

{\bf Model 75}: For $\mu \sim {\cal O}(200)$ GeV, $m_{\chi_1^+}-m_{\chi_1^0}$ 
and $m_{\chi_2^0} - m_{\chi_1^0}$ are very small even after the inclusion of 
the radiative corrections [Fig.~1] and the decays of $\chi_2^0$ and 
$\chi_1^+$ into the LSP and massive fermions are not 
kinematically possible [in this case, these particles if they are not almost 
stable, will decay into the LSP and soft pions]. For large values of $\mu$, 
the mass differences between $\chi_1^+, \chi_2^0$ and the LSP are sizeable 
[although penalizing the $b\bar{b}$ final state of $\chi_2^0$] and the 
different evolution of the sfermion masses as a function of the gaugino masses 
and the different radiative corrections to the bottom and tau lepton masses, 
explain the quantitative  differences between the branching ratios in the 
two models {\bf 75} and {\bf 1}. \s

{\bf Model 200}: Here, the chargino $\chi_1^+$ and the LSP are wino--like for
large values of $\mu$, and the mass difference $m_{\chi_1^+}-m_{\chi_1^0}$ is
too small for the decay $\chi_1^+ \to \chi_1^0 \tau^+ \nu$ to occur. For
smaller $\mu$ values, this decay can receive large contributions from light
Higgs bosons and sizeable ones from light sfermions [in particular,
$\tilde{b}_1$ is lighter than in model {\bf 1}]. In the case of the decays of
the neutralino $\chi_2^0$, since the difference $m_{\chi_2^0}- m_{\chi_1^0}$ is
always large [exceeding $M_Z$ for $\mu \gsim 200$ GeV i.e.  when $\chi_2^0$ is
bino--like] the branching ratios BR$(\chi_2^0 \to \chi_1^0 \tau^+ \tau^- ,
\chi_1^0 b\bar{b}$) are similar to those of model {\bf 24} and are controlled by
the $Z$ boson decay branching ratios. Note that in this scenario, the decay
$\chi_2^0 \to \tilde{g} q\bar{q}$ is possible as will be discussed later. \s

{\bf Model OII}: In this model, the situation is similar to model {\bf 200}
for the decays of $\chi_1^+$. Indeed, for $\mu \gsim 300$ GeV, $\chi_1^+$
is almost degenerate with the LSP and the channel $\chi_1^+ \to \chi_1^0 \tau^+
\nu$ is kinematically closed. This is almost the case for the neutralino
$\chi_2^0$ which has a mass that is close to the LSP mass for large
$\mu$ values, suppressing the $b\bar{b}$ decay mode. However, the new feature
in this scenario  is that $M_3 <M_{1,2}$ and for large $\mu$ values, the decay 
modes $\chi_1^+, \chi_2^0, \chi_1^0 \to \tilde{g} q\bar{q}$ open up and become 
dominant because of the strong interaction part [note, however, that the 
neutralino $\chi_1^0$ is not the LSP anymore]. \smallskip

\begin{figure}[htbp]
\vspace*{-4.5cm}
\hspace*{-2.3cm}
\mbox{\psfig{figure=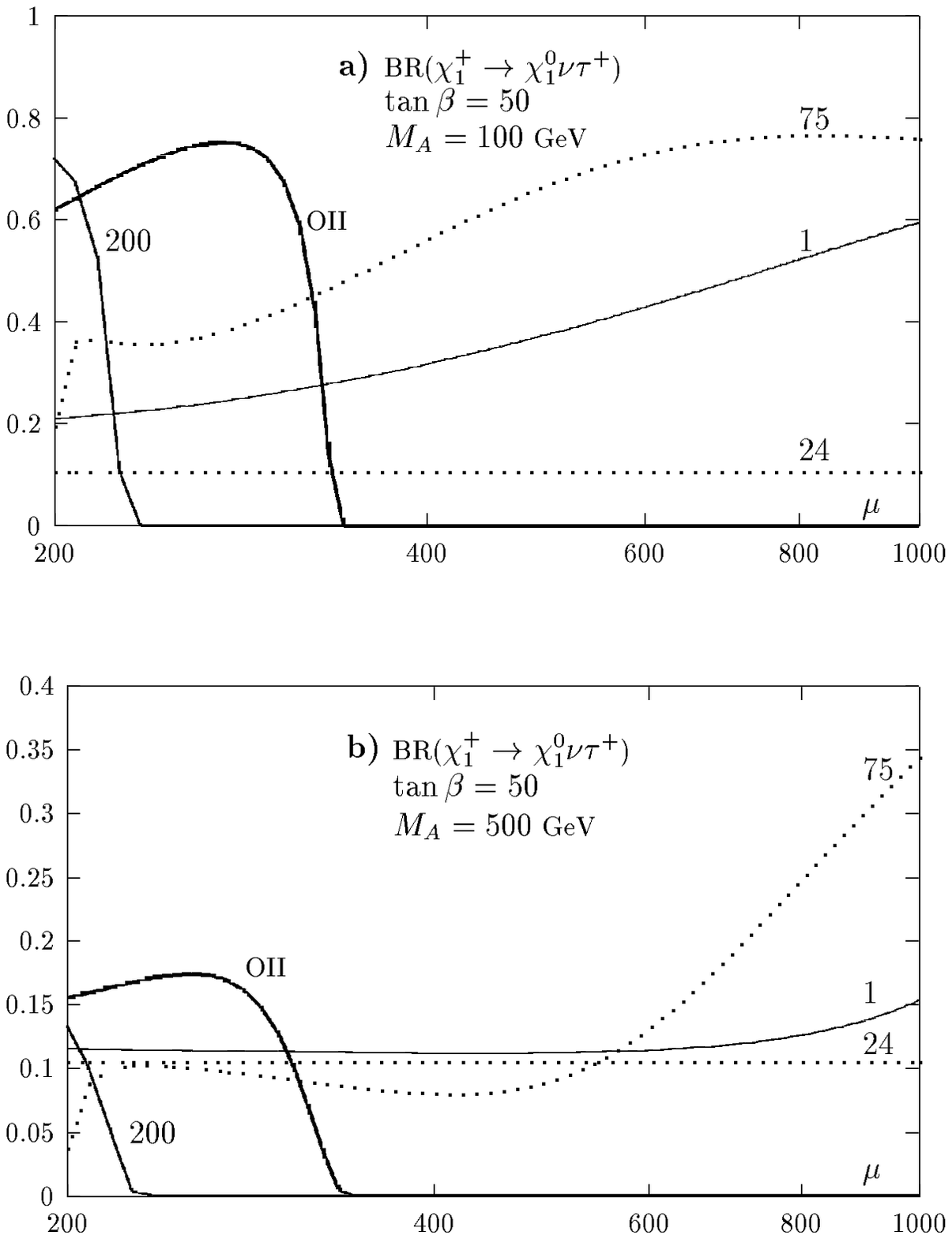,width=22.5cm}}
\vspace*{-9.cm}
\caption[]{\it The branching ratios BR$(\chi^+_1 \ra \chi_1^0 \nu \tau^+)$ 
as a function of $\mu$ in models with non--universal gaugino masses; we have
fixed the parameters to $\tb=50$, $m_0=500$ GeV, $M_2=150$ GeV and $M_A=100$ 
$(500)$ GeV for a (b).} 
\end{figure}

\begin{figure}[htbp]
\vspace*{-4.5cm}
\hspace*{-2.3cm}
\mbox{\psfig{figure=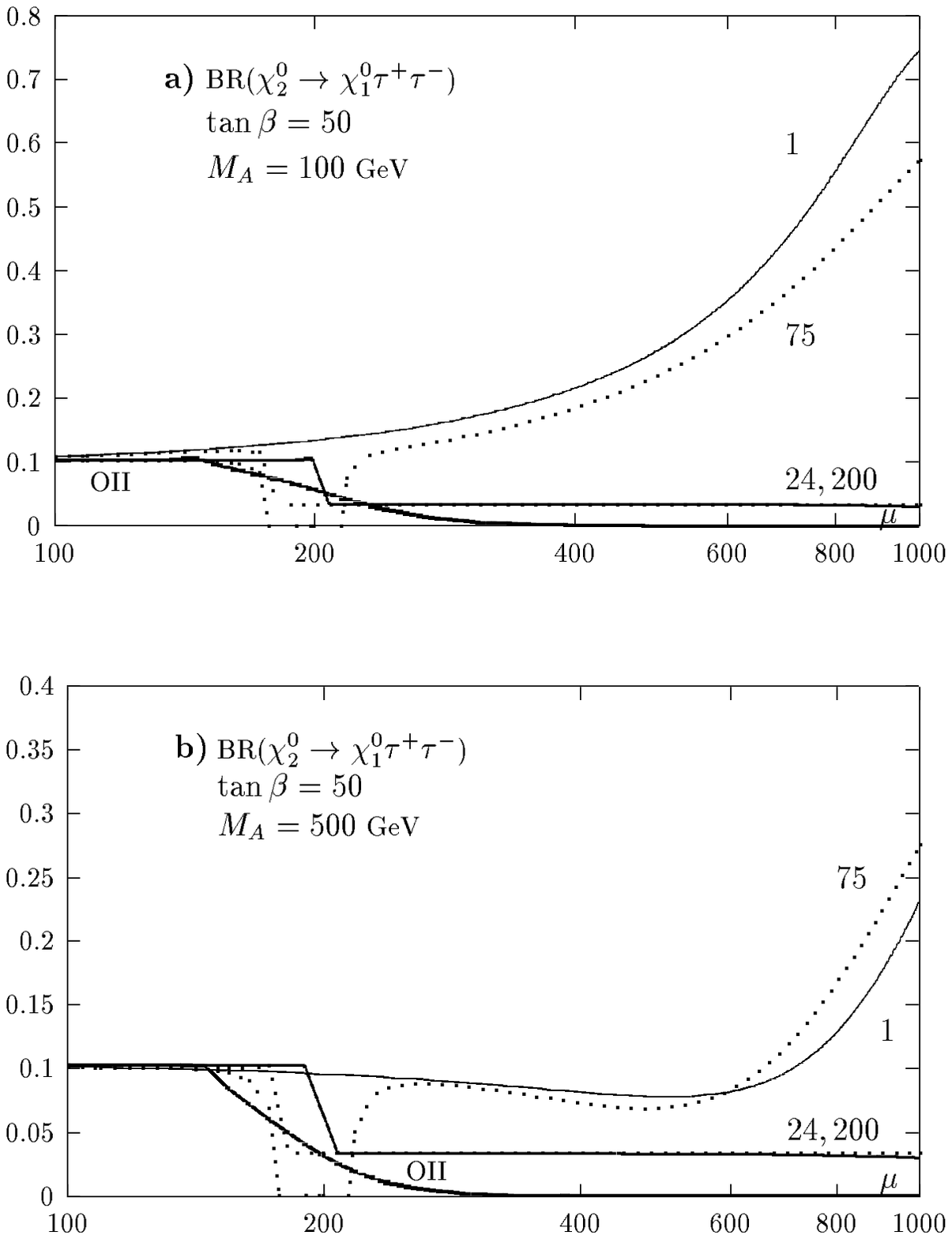,width=22.5cm}}
\vspace*{-9.cm}
\caption[]{\it The branching ratios BR$(\chi^0_2 \ra \chi_1^0 \tau^+ \tau^-)$ 
as a function of $\mu$ in models with non--universal gaugino masses; we have
fixed the parameters to $\tb=50$, $m_0=500$ GeV, $M_2=150$ GeV and $M_A=100$ 
$(500)$ GeV for a (b).} 
\end{figure}

\begin{figure}[htbp]
\vspace*{-4.5cm}
\hspace*{-2.3cm}
\mbox{\psfig{figure=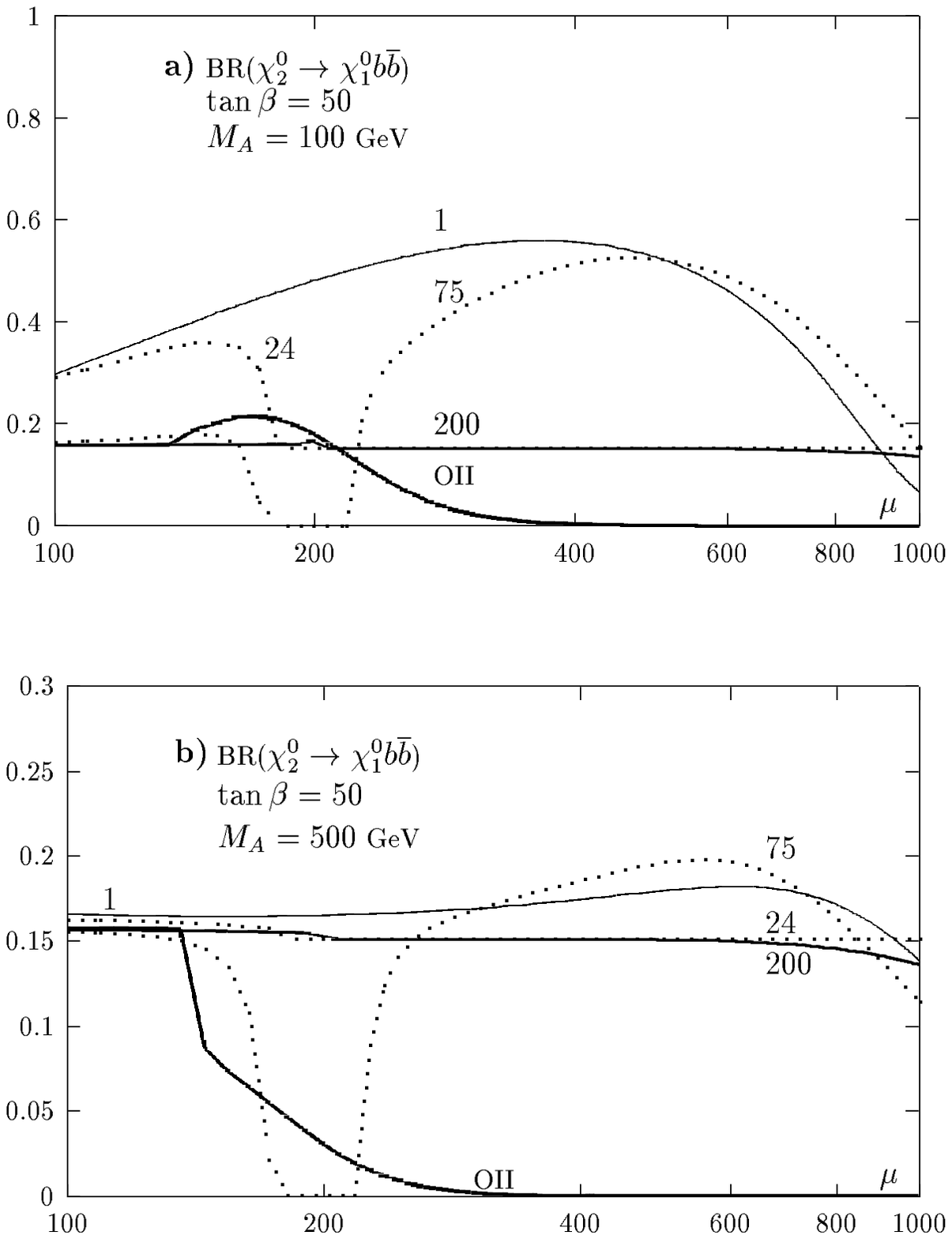,width=22.5cm}}
\vspace*{-9.cm}
\caption[]{\it The branching ratios BR$(\chi^0_2 \ra \chi_1^0 b \bar{b})$ 
as a function of $\mu$ in models with non--universal gaugino masses; we have
fixed the parameters to $\tb=50$, $m_0=500$ GeV, $M_2=150$ GeV and $M_A=100$ 
$(500)$ GeV for a (b).} 
\end{figure}

Finally, Fig.~13 shows the branching ratio for the decays $\chi_2^0 \to
\tilde{g} q\bar{q}$ in the model {\bf 200} where $m_{\chi_1^0} < m_{\tilde{g}}
<m_{\chi_2^0}$. For small $\mu$ values, the lightest neutralinos are
higgsino--like and they are degenerate in mass. For values of $\mu$ around
$M_2$, the hierarchy $m_{\chi_1^0} < m_{\tilde{g}} < m_{\chi_2^0}$ is possible
while $\chi_1^0$is the LSP, and the decay can occur. However, the neutralino
couplings to quark--squark pairs are small except in the case of (s)bottoms for
large $\tb$ values. In contrast, the $\chi_1^0$--$\chi_2^0$--$Z$ coupling is
almost maximal here. BR($\chi_2^0 \to \tilde{g} \sum q\bar{q}) $, which is
approximately the same as BR($\chi_2^0 \to \tilde{g}b\bar{b})$, is thus not
dominant, but can reach the level of 25\%, despite of the fact that it is a
mixed strong--electroweak decay mode.  For larger values of $\mu$, the
neutralinos $\chi_{1,2}^0$ become gaugino--like and the partial decay widths
$\Gamma(\chi_2^0 \to \tilde{g} q\bar{q})$ are more important since the
couplings to fermion--sfermion pairs are enhanced; however in this case,
because $M_3 < M_2$, the gluino becomes lighter than the lightest neutralino
which we assume here to be the LSP. \s

In the case of the charginos, the branching ratio for the decays $\chi_1^+ \to
\tilde{g} q\bar{q}'$ for higgsino--like charginos is even smaller, since there 
is no final state with massive fermions [the $t\bar{b}$ decay mode is not 
kinematically accessible] and the first and second generation (s)particles have 
small couplings for higgsino--like charginos. In the gaugino--like region, 
the lightest chargino becomes lighter than the gluino and the decay does 
not occur. \bigskip

\begin{figure}[htbp]
\vspace*{-4.5cm}
\hspace*{-2.3cm}
\mbox{\psfig{figure=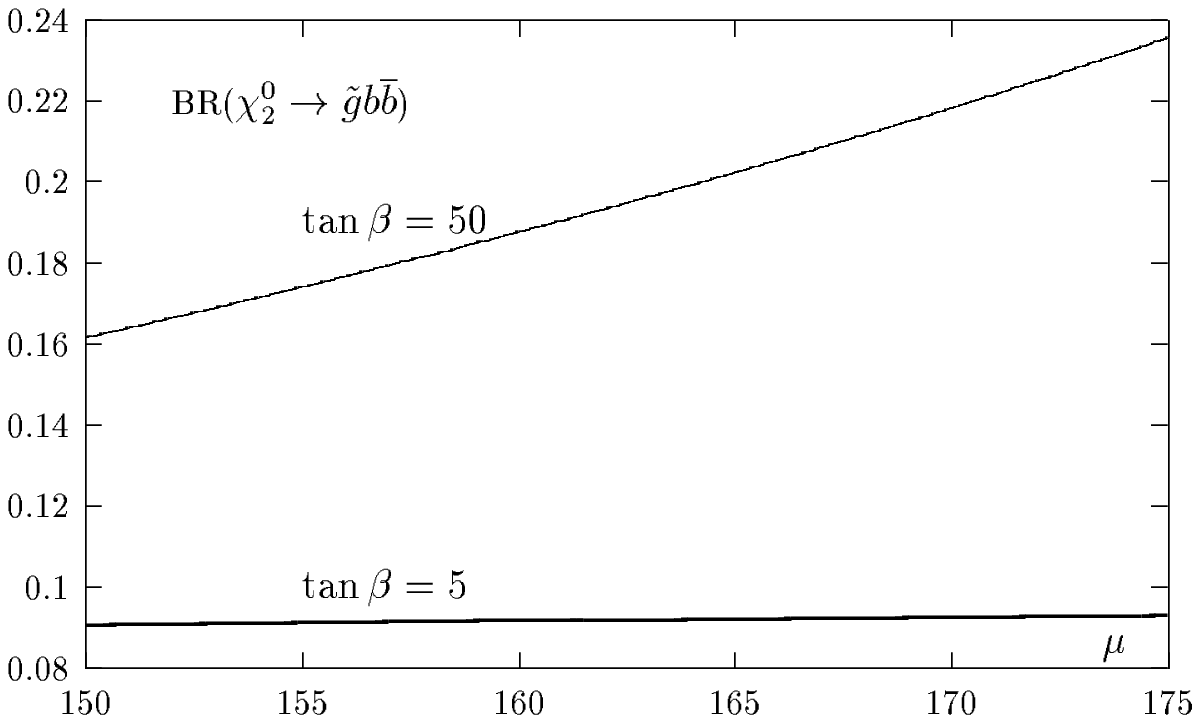,width=22.5cm}}
\vspace*{-20.cm}
\caption[]{\it The branching ratio BR$(\chi^0_2 \ra \tilde{g} q \bar{q})$ 
as a function of $\mu$ in model {\bf 200} with non--universal gaugino masses.
The parameters are fixed to: $\tb=5$ and 50, $M_2=150$ GeV, $m_0= M_A=500$ 
GeV.} 
\end{figure}

\newpage

We have developed a fortran code called {\tt SDECAY} \cite{sdecay} which
calculates the partial decay widths and branching ratios of the chargino and
neutralino decays. It includes not only the three--body decays,
$\chi_2^0 \to \chi_1^0 f \bar{f}$ and $\chi_1^+ \to \chi_1^0 f \bar{f}'$
discussed in this paper, but also all the two--body decays of the charginos
and neutralinos [including the heavy $\chi_{3,4}^0$ and $\chi_2^+$ states]
into gauge bosons, MSSM Higgs bosons and fermion--sfermion pairs. The program 
contains, in addition, the branching ratios for the two--, three-- and 
four--body decay modes of the top squarks, as well as the three--body decays of gluinos and all relevant decay modes of sfermions other than the top squarks. 
\s

The gaugino mass parameters $M_1,M_2,M_3$, as well as the soft--SUSY breaking
scalar masses $m_{\tilde{f}_L}$ and $m_{\tilde{f}_R}$, can be chosen as free
parameters so that decay widths and branching ratios can be obtained in
non--universal models. However, scenarios with boundary conditions at high
scales are also implemented, since the program has been interfaced with the
code {\tt SUSPECT} \cite{suspect} for the the renormalization group equations
for parameter evolution and for the proper breaking of the electroweak
symmetry. For the parameterization of the MSSM Higgs sector, the program  has 
been interfaced with the code {\tt HDECAY} \cite{hdecay}, which in addition 
gives the decay products for the Higgs particles. All radiative corrections 
discussed in this analysis are incorporated into the program. \s

We have compared our results with those of
Ref.~\cite{R4a} which have been implemented in the program {\tt ISAJET}
\cite{ISAJET}. For massless fermions and if the SUSY radiative correction to
the fermion masses are not taken into account in the sfermion mass matrices,
the agreement was very good in models with gaugino mass unification, giving a
great confidence that this rather involved calculation is correct. [The
comparison was slightly involved since the evolution of the couplings and the
soft SUSY--breaking terms as well as the parameterization of the Higgs sectors
are given in different approximations in the programs {\tt SUSPECT} and {\tt
ISAJET} and we needed to use the same input parameters at low energy in both
programs\footnote{We thank Laurent Duflot from ALEPH for his help with this
comparison.  An independent numerical check in the case of the chargino decays 
into massless final state fermions, has also been performed by F. Boudjema and 
V. Lafage \cite{FB}.}.] Our results are however different from those which can 
be obtained with the program {\tt SUSYGEN} (version 2.2) \cite{SUSYGEN} used for
SUSY particle searches at LEP, since in the latter version, the Higgs boson
exchange contributions and the effect of third generation sfermion mixing have
not been implemented\footnote{These effects are being included in a new version
of the program; we thank S. Katsanevas and N. Ghodbane for discussions on this
issue.}.  

\clearpage
\newpage

\section*{5. Conclusions}

In this paper, we have analyzed the decay modes of charginos and neutralinos in
the MSSM where the lightest neutralino $\chi_1^0$ is the LSP. We focused on the
three--body decay modes of the lightest charginos $\chi_1^\pm$ and the
next--to--lightest neutralinos $\chi_2^0$ into the LSP and two fermion final
states, and made a complete calculation of the decay widths and branching
ratios, taking into account all possible channels: vector boson, Higgs boson
and sfermion exchange with the mixing in the sfermion sector included.  In this
context, we have shown that the SUSY radiative corrections to the heavy fermion
masses, in particular to the $b$--quark mass, and to the chargino and
neutralino masses can play an important role.  We derived full analytical
expressions of the Dalitz densities and the integrated partial decay widths in
the massless fermion case, and provided the complete formulae for the
differential decay widths, including the finite masses of the final fermions
and the polarization of the decaying charginos or neutralinos.  A fortran code
for the numerical evaluation of all the branching ratios is made available 
\cite{these}. \s

For large values of $\tan \beta$, the bottom and tau Yukawa couplings become
large, leading to smaller masses of the tau slepton and bottom squark compared
to their first and second generation partners. At the same time, the Yukawa
couplings of tau and bottom quarks to the Higgs bosons can become very large. 
The branching ratios of the decays of the lightest chargino into $\tau \nu$
final states and of the next--to--lightest neutralino into $b\bar{b}$ and
$\tau^+ \tau^-$ pairs can be thus strongly enhanced in this scenario. We have
illustrated this possibility in mSUGRA--type scenarios where the gaugino masses
are unified at the GUT scale, but also in scenarios where the boundary
conditions for binos and winos are different at this high scale, leading to
different mass patterns for the charginos and neutralinos, which affect the
decay branching ratios. In particular, new decay channels, such as the decay 
of the lightest chargino and the next--to--lightest neutralino into gluino and 
quark--antiquark final states, open up kinematically and can play an important 
role. \s

When SUSY particles will decay via cascades through charginos and the heavier
neutralinos, the events will contain more $\tau$ leptons and $b$--quarks, than
first and second generation leptons and quarks. This renders the search for
SUSY particles and the measurement of the SUSY parameters, where the electron
and muon channels where used, less straightforward as already discussed in
Ref.~\cite{R4b}.  $b$--tagging and the identification of the decays of the
tau leptons become then a crucial issue in the search and the study of the
properties of these particles, in particular at hadron colliders such as the
Tevatron and LHC.  

\bigskip 

\nn {\bf Acknowledgments:} \s

\nn We thank the members of the French GDR--Supersym\'etrie, in particular 
F.~Boudjema, D.~Denegri, L.~Duflot, J-F.~Grivaz, S.~Katsanevas, J-L.~Kneur, 
V.~Lafage and S. Rosier--Lees, for various discussions. Thanks also go to 
Asesh Datta, Jean-Loic Kneur and Michael Spira for discussions on the clashing 
arrows of Majorana particles, and for JLK for his additional help with the 
Fortran code. This work is partially supported by the European Union under 
contract HPRN-CT-2000-00149. Y.M is supported by a MNERT fellowship and thanks 
the members of the SPN for the facilities accorded to him. 

\newpage
\setcounter{equation}{0}
\renewcommand{\theequation}{A.\arabic{equation}}
\section*{Appendix}

In this Appendix, we will give the lengthy formulae for the three--body partial
decay widths in the case of finite masses for the fermion final states $[\mu_u
\neq \mu_d \neq 0$, with $\mu_f=m_f^2/m_{\chi_i}^2$] and where the polarization
of the decaying chargino or neutralino is taken into account\footnote{The
expressions for three--body decays of charginos and neutralinos, including the
polarization of the initial states, are available in the literature, see
Ref.~\cite{pol}, in the case of massless final fermions, no Higgs boson
exchange and no mixing in the sfermion sector.}:
\beq 
\chi_i (q, n_{\chi_i}) \, \rightarrow \, \chi_j^0 (p) \ u (p_1) \ \bar{d} (p_2) 
\eeq
where $q, p, p_1$ and $p_2$ are the four--momenta of the particles and
$n_{\chi_i}$ is the spin four--vector of the decaying ``ino" defined by 
$n_{\chi_i} \cdot n_{\chi_i} =-1$ and $n_{\chi_i} \cdot q =0$. \s
 
The partial decay width for both chargino and neutralino three--body decays, 
following the notation given in section 3, is given  by: 
\beq
\frac{ \dx \Gamma_{\chi_i} } {\dx \hat u \dx \hat t}  &=& 
\frac{e^4 m_{\chi_i}} 
{64(2\pi)^3} \, N_c \\ && \bigg[ \dx \Gamma_V + \dx \Gamma_{\tilde{u}}
+ \dx \Gamma_{\tilde{d}} + \dx \Gamma_{\Phi} +\dx \Gamma_{H_1 H_2} + \dx 
\Gamma_{V \tilde{u}} +  
\dx \Gamma_{V \tilde{d}} + \dx \Gamma_{\tilde{u} \tilde{d}} + \dx \Gamma_{\Phi 
\tilde{u}} + \dx \Gamma_{\Phi \tilde{d}}  \bigg] \non 
\eeq
where $\dx \Gamma_X$ is decomposed into the spin--independent part [which is
half of the unpolarized partial decay width] and the part which depends on
the spin four vector of the decaying ``ino": 
\beq
\dx \Gamma_X = \frac{1}{2}\dx \Gamma^U_X + \dx \Gamma^S_X
\eeq
We will use the reduced Mandelstam variables and spin vector:
\beq
\hat u = (q-p_1)^2/m_{\chi_i}^2 \quad , \qquad \hat t = (q-p_2)^2/m_{\chi_i}^2
\quad {\rm and}  \qquad n= n_{\chi_i}/m_{\chi_i}
\eeq

\subsection*{Spin--independent part:}

\beq
\dx \Gamma^U_{V} &=& \frac{4}{(1+\mu_\chi + \mu_u + \mu_d - \mu_V - 
\hat u-\hat t)^2} 
\Bigg\{
\left[ ( G^L_{jiV} )^2 + ( G^R_{jiV} )^2 \right] 
\bigg( 
\left[ ( v_V^f )^2 + ( a_V^f )^2 \right]
\non\\
&&
[ (1+\mu_\chi +\mu_u +\mu_d ) (\hat u+\hat t)- \hat u^2-\hat t^2 -
2\mu_u \mu_d - \mu_u - \mu_d -\mu_\chi (2+\mu_u+\mu_d) ] \non\\
&& + 2\left[ ( v_V^f )^2 - ( a_V^f )^2 \right]
\sqrt{\mu_u \mu_d} [\hat u+\hat t - \mu_u-\mu_d] \bigg) 
+ 2 \left[ ( G^L_{jiV} )^2 - ( G^R_{jiV} )^2 \right] v_V^f a_V^f
\non\\
&& [ \hat u^2-\hat t^2 - (\hat u-\hat t) (1+\mu_\chi+\mu_u+\mu_d)
+ (\mu_d-\mu_u)(1-\mu_\chi) ] + 4 G^L_{jiV} G^R_{jiV}\sqrt{\mu_\chi}
\non\\
&& \bigg(\left[ ( v_V^f )^2 + ( a_V^f )^2 \right] [\hat u+\hat t
-1-\mu_\chi)] - 4 \left[ ( v_V^f )^2 - ( a_V^f )^2 \right] \sqrt{\mu_u \mu_d}
\bigg) \Bigg\}
\eeq
\newpage
\beq
\dx \Gamma^U_{\Phi} &=& \sum_k\frac{2}{(1+\mu_\chi + \mu_u + \mu_d - \mu_k - 
\hat u-\hat t)^2} 
\Bigg\{ 
\left[ ( G^L_{ijk} )^2 + ( G^R_{ijk} )^2 \right] 
\bigg( 
\left[ ( v_k^f )^2 + ( a_k^f )^2 \right] 
\non\\
&& 
[ (1+\mu_\chi +\mu_u +\mu_d ) (\hat u+\hat t) - 
(\hat u+\hat t)^2 - (1+\mu_\chi)(\mu_u+\mu_d) ] 
-2\left[ ( v_k^f )^2 - ( a_k^f )^2 \right]
\non\\
&& \sqrt{\mu_u\mu_d} [ \hat u+\hat t - \mu_u -\mu_d] \bigg) + 
4 G^L_{ijk} G^R_{ijk} \sqrt{\mu_\chi}\bigg( 
\left[ ( v_k^f )^2 + ( a_k^f )^2 \right][ 1+\mu_\chi-\hat u-\hat t ]
\non\\
&&  
 -2 \left[ ( v_k^f )^2 - ( a_k^f )^2 \right] \sqrt{\mu_u\mu_d} \bigg) 
\Bigg\}
\eeq
\beq
\dx \Gamma^U_{H_1 H_2} &=& \frac{4 v_{H_1}^f v_{H_2}^f}
{(1+\mu_\chi + \mu_u + \mu_d - \mu_{H_1}- \hat u-\hat t)
(1+\mu_\chi + \mu_u + \mu_d - \mu_{H_2} - \hat u-\hat t)} \non\\
&& \Bigg\{
2\mu_\chi \left[G^L_{ij1}G^R_{ij2}+G^L_{ij2}G^R_{ij1} \right] 
[1+\mu_\chi-2\sqrt{\mu_u\mu_d}-\hat u-\hat t] + 
\left[ G^L_{ij1}G^L_{ij2}+G^R_{ij1}G^R_{ij2} \right] 
\non\\
&& [(\mu_u+\mu_d)(-1+2\sqrt{\mu_u\mu_d}
-\mu_\chi)+(\hat u+\hat t)(1+\mu_\chi+\mu_u+\mu_d-2\sqrt{\mu_u \mu_d})
\non\\
&& - (\hat u+\hat t)^2 ] \Bigg\}
\eeq
\beq
\dx \Gamma^U_{V\Phi} &=& \sum_{k=1}^2 \frac{8}
{(1+\mu_\chi + \mu_u + \mu_d - \mu_{H_k}- \hat u-\hat t)
(1+\mu_\chi + \mu_u + \mu_d - \mu_{V} - \hat u-\hat t)} \non\\
&& \Bigg\{
\left[ G^L_{jiV} G^R_{ijk} + G^R_{jiV} G^L_{ijk}\right]
\Bigg( \left[ v_k^f v_V^f + a_k^f a_V^f \right]
\sqrt{\mu_u} (-\mu_\chi-\mu_d+\hat u) + \left[ v_k^f v_V^f - a_k^f a_V^f\right]
\non\\
&& 
\sqrt{\mu_d} (\mu_u + \mu_\chi-\hat t) \Bigg) + 
\left[ G^L_{jiV} G^L_{ijk} + G^R_{jiV} G^R_{ijk} \right] 
\Bigg( \left[ v_k^f v_V^f + a_k^f a_V^f \right]
\sqrt{\mu_u \mu_\chi} (1+\mu_d-\hat t)
\non\\
&& + \left[ v_k^f v_V^f - a_k^f a_V^f \right]
\sqrt{\mu_d \mu_\chi} (-1-\mu_u+\hat u) \Bigg) \Bigg\}
\eeq
\beq
\dx \Gamma^U_{\tilde u} &=& \sum_{k,l=1}^2 \frac{1}
{ (- \mu_d - \mu_{\tilde u_k}+ \hat t)
(- \mu_d - \mu_{\tilde u_l} + \hat t)} \Bigg\{
-4 a_1^u \sqrt{\mu_\chi} \sqrt{\mu_u\mu_d} + 2 a_2^u \sqrt{\mu_\chi \mu_u}
\non \\ 
&& (-\mu_d -1 +\hat t) + 2 a_3^u \sqrt{\mu_d} (-\mu_u-\mu_\chi+\hat t)
 + a_4^u [-\hat t^2 + \hat t (1+\mu_\chi+\mu_d+\mu_u) 
\non\\
&& - (\mu_\chi+\mu_u)(1+\mu_d)] \Bigg\}
\eeq
\beq
\dx \Gamma^U_{\tilde d} &=& \sum_{k,l=1}^2 \frac{1}
{(- \mu_u - \mu_{\tilde d_k}+ \hat u)
(- \mu_u - \mu_{\tilde d_l} + \hat u)} \Bigg\{
-4 a_1^d \sqrt{\mu_\chi} \sqrt{\mu_u\mu_d} + 2 a_2^d \sqrt{\mu_\chi \mu_d}
\non \\ 
&& (-\mu_u-1 +\hat u) + 2 a_3^d \sqrt{\mu_u} (-\mu_d-\mu_\chi+\hat u)
 + a_4^d [-\hat u^2 + \hat u (1+\mu_\chi+\mu_d+\mu_u) 
\non\\
&& - (\mu_\chi+\mu_d)(1+\mu_u)] \Bigg\}
\eeq
where
\beq
a_1^f &=& (a^f_{jk} b^f_{jl} + a^f_{jl} b^f_{jk}) 
(a^f_{ik} b^f_{il} + a^f_{il} b^f_{ik}) \non\\
a_2^f &=& (a^f_{jk} b^f_{jl} + a^f_{jl} b^f_{jk})
(a^f_{ik} a^f_{il} + b^f_{ik} b^f_{il}) \non\\
a_3^f &=& (a^f_{jk} a^f_{jl} + b^f_{jk} b^f_{jl})
(a^f_{ik} b^f_{il} + a^f_{il} b^f_{ik}) \non\\
a_4^f &=& (a^f_{jk} a^f_{jl} + b^f_{jk} b^f_{jl})
(a^f_{ik} a^f_{il} + b^f_{ik} b^f_{il})
\eeq
\beq
\dx \Gamma^U_{V\tilde d} &=& \sum_{l=1}^2 \frac{-4}
{(1+\mu_\chi + \mu_u + \mu_d - \mu_{V}- \hat u-\hat t)
(- \mu_u - \mu_{\tilde d_l} + \hat u)} \Bigg\{ b_1^{f}(jiV) [
-(\mu_\chi+\mu_d)(\mu_u+1)\ \non\\
&& -\hat u^2 + \hat u(1+\mu_u+\mu_d+\mu_\chi)] + b_2^{f}(jiV) \sqrt{\mu_\chi}
(\hat u+\hat t-1-\mu_\chi) + b_3^{f}(jiV) \sqrt{\mu_u\mu_d}  
\non\\
&& (\hat u+\hat t -\mu_u-\mu_d) - 4 b_4^{f}(jiV) \sqrt{\mu_\chi} 
\sqrt{\mu_u\mu_d} + 
b_5^{f}(jiV) \sqrt{\mu_d} (\hat t-\mu_\chi-\mu_u) 
\non\\
&& +b_6^{f}(jiV)\sqrt{\mu_u\mu_\chi} (\hat t-\mu_d-1) 
+ 2b_7^{f}(jiV) \sqrt{\mu_u} (\hat u-\mu_\chi-\mu_d) + 2b_8^{f}(jiV) 
\sqrt{\mu_\chi\mu_d} \non\\
&&  (\hat u-\mu_u-1) \Bigg\}
\eeq
\beq
\dx \Gamma^U_{V\tilde u} &=& \sum_{l=1}^2 \frac{4}
{(1+\mu_\chi + \mu_u + \mu_d - \mu_{V}- \hat u-\hat t)
(- \mu_d - \mu_{\tilde u_l} + \hat t)} \Bigg\{ b_1^{f}(jiV) \sqrt{\mu_\chi}
(\hat u+\hat t-1  \non\\
&& -\mu_\chi) +b_2^{f}(jiV) [-\hat t^2 + \hat t(1+\mu_u + \mu_d+\mu_\chi)-
(\mu_\chi+\mu_u)(\mu_d+1)] - 4 b_3^{f}(jiV) 
\non\\
&& \sqrt{\mu_\chi} \sqrt{\mu_u\mu_d}
+ b_4^{f}(jiV) \sqrt{\mu_u\mu_d} (\hat u+\hat t -\mu_u-\mu_d)  
+ 2 b_5^{f}(jiV) \sqrt{\mu_u\mu_\chi} (\hat t-\mu_d-1)   \non\\
&& + 2 b_6^{f}(jiV)
\sqrt{\mu_d} (\hat t-\mu_\chi-\mu_u) 
+ b_7^{V}(jiV) \sqrt{\mu_\chi\mu_d} (\hat u-\mu_u-1) + 
b_8^{f}(jiV) \sqrt{\mu_u} 
\non\\
&& (\hat u-\mu_\chi-\mu_d) \Bigg\}
\eeq
where
\beq
b_1^{f}(ijk) = a^f_{il} a^f_{jl} G^R_{ijk} (v_k^f + a_k^f) + b^f_{il} b^f_{jl} 
G^L_{ijk} (v_k^f - a_k^f) \non\\
b_2^{f}(ijk) = a^f_{il} a^f_{jl} G^L_{ijk} (v_k^f + a_k^f) + b^f_{il} b^f_{jl} 
G^R_{ijk} (v_k^f - a_k^f) \non\\
b_3^{f}(ijk) = a^f_{il} a^f_{jl} G^R_{ijk} (v_k^f - a_k^f) + b^f_{il} b^f_{jl} 
G^L_{ijk} (v_k^f + a_k^f) \non\\
b_4^{f}(ijk) = a^f_{il} a^f_{jl} G^L_{ijk} (v_k^f - a_k^f) + b^f_{il} b^f_{jl} 
G^R_{ijk} (v_k^f + a_k^f) \non\\
b_5^{f}(ijk) = a^f_{jl} b^f_{il} G^L_{ijk} (v_k^f + a_k^f) + a^f_{il} b^f_{jl}
G^R_{ijk} (v_k^f - a_k^f) \non\\
b_6^{f}(ijk) = a^f_{jl} b^f_{il} G^R_{ijk} (v_k^f - a_k^f) + a^f_{il} b^f_{jl}
G^L_{ijk} (v_k^f + a_k^f) \non\\
b_7^{f}(ijk) = a^f_{jl} b^f_{il} G^L_{ijk} (v_k^f - a_k^f) + a^f_{il} b^f_{jl}
G^R_{ijk} (v_k^f + a_k^f) \non\\
b_8^{f}(ijk) = a^f_{jl} b^f_{il} G^R_{ijk} (v_k^f + a_k^f) + a^f_{il} b^f_{jl}
G^L_{ijk} (v_k^f - a_k^f) \non\\
\eeq
\newpage
\beq
\dx \Gamma^U_{\tilde u\tilde d} &=& \sum_{k,l=1}^2 \frac{-2}
{(- \mu_u - \mu_{\tilde d_l} + \hat u)
(- \mu_d - \mu_{\tilde u_k} + \hat t)} \Bigg\{ [a^u_{ik} a^u_{jk} b^d_{il} 
b^d_{jl} + a^d_{il} a^d_{jl} b^u_{ik} b^u_{jk}] \sqrt{\mu_u\mu_d} 
 \non\\ 
&& (\hat u+\hat t-\mu_u-\mu_d) 
+ [a^u_{ik} a^u_{jk} a^d_{il} b^d_{jl} + b^u_{ik} b^u_{jk} 
b^d_{il} a^d_{jl}] \sqrt{\mu_d} (\hat t-\mu_\chi-\mu_u)
\non\\
&& + [a^u_{jk}a^d_{il} a^d_{jl}b^u_{ik} + b^u_{jk}b^d_{il} b^d_{jl}a^u_{ik}]
\sqrt{\mu_\chi\mu_d}(\hat u-\mu_u-1) -2[a^u_{jk} a^d_{jl} b^u_{ik} b^d_{il} 
+ a^u_{ik} a^d_{il} b^u_{jk} b^d_{jl}] 
\non\\
&& \sqrt{\mu_\chi} \sqrt{\mu_u\mu_d} +
[a^u_{jk} b^u_{ik} b^d_{il} b^d_{jl} 
+ a^u_{ik} a^d_{il} a^d_{jl} b^u_{jk}] \sqrt{\mu_u} (\hat u-\mu_\chi-\mu_d)
 \non\\
&& + [a^u_{jk} a^d_{il} b^u_{ik} b^d_{jl} + a^u_{ik} a^d_{jl}
 b^u_{jk} b^d_{il}]
(\hat u \hat t -\mu_\chi -\mu_u \mu_d) +
[a^u_{ik} a^u_{jk} a^d_{il} a^d_{jl} + b^u_{ik} b^u_{jk} b^d_{il} b^d_{jl}] 
\sqrt{\mu_\chi} 
\non\\
&& (\hat u+\hat t-\mu_\chi-1) + [a^u_{ik} a^u_{jk} a^d_{jl} b^d_{il} 
+ b^u_{ik}b^u_{jk} b^d_{jl} a^d_{il}] \sqrt{\mu_\chi\mu_u} 
(\hat t-\mu_d-1) \Bigg\}
\eeq
\beq
\dx \Gamma^U_{\Phi_k \tilde d} &=& \sum_{k,l} \frac{2}
{(1+\mu_\chi + \mu_u + \mu_d - \mu_{\Phi_k}- \hat u-\hat t)
(- \mu_u - \mu_{\tilde d_l} + \hat u)} \Bigg\{ b_1^d(ijk)\sqrt{\mu_\chi\mu_u}
\non\\
&&(\hat t-\mu_d-1)
+ b_2^d(ijk)\sqrt{\mu_u} (-\hat u+\mu_\chi+\mu_d) + b_3^d(ijk) 
\sqrt{\mu_\chi\mu_d} (-\hat u+\mu_u+1)  \non\\
&&+ b_4^d(ijk) \sqrt{\mu_d} (\hat t -\mu_\chi-\mu_u)
+ b_5^d(ijk) [\hat u^2+\hat u \hat t -\hat u(1+\mu_u+\mu_d+\mu_\chi)
+\mu_u\mu_\chi + \mu_d]\non\\
&& 
+ 2 b_6^d(ijk) \sqrt{\mu_\chi} \sqrt{\mu_u\mu_d}
+ b_7^d(ijk) \sqrt{\mu_u\mu_d} (\hat u+\hat t-\mu_u -\mu_d)
\non\\
&& + b_8^d(ijk) \sqrt{\mu_\chi} (\hat u+\hat t-\mu_\chi-1) \Bigg\}
\eeq
\beq
\dx \Gamma^U_{\Phi_k \tilde u} &=& \sum_{k,l} \frac{2}
{(1+\mu_\chi + \mu_u + \mu_d - \mu_{\Phi_k}- \hat u-\hat t)
(- \mu_d - \mu_{\tilde u_l} + \hat t)} \Bigg\{ b_1^u(ijk)\sqrt{\mu_u}
(\hat u-\mu_\chi-\mu_d) \non\\
&&+ b_2^u(ijk)\sqrt{\mu_\chi\mu_u} (-\hat t+1+\mu_d) + b_3^u(ijk) 
\sqrt{\mu_d} (-\hat t+\mu_u+\mu_\chi) + b_4^u(ijk) \sqrt{\mu_\chi\mu_d} 
\non\\
&& (\hat u-1-\mu_u) +2 b_5^u(ijk)\sqrt{\mu_\chi}\sqrt{\mu_u\mu_d}
+ b_6^u(ijk) [\hat u \hat t + \hat t^2 - 
\hat t(1+\mu_u+\mu_d+\mu_\chi)+\mu_u 
\non\\
&& + \mu_\chi \mu_d] + b_7^u(ijk) \sqrt{\mu_\chi} (\hat u+\hat t-\mu_\chi -1)
+ b_8^u(ijk) \sqrt{\mu_u\mu_d} 
(\hat u+\hat t-\mu_u-\mu_d) \Bigg\}
\eeq

\subsection*{Spin--dependent part:}
\beq
\dx \Gamma^S_{V} &=& 
\frac{4}{(1+\mu_\chi + \mu_u + \mu_d - \mu_V - \hat u-\hat t)^2} 
\Bigg\{
\left[ ( G^L_{jiV} )^2 - ( G^R_{jiV} )^2 \right] 
\bigg( 
\left[ ( v_V^f )^2 + ( a_V^f )^2 \right]
\non\\ 
&& [p_1.n(\mu_\chi+\mu_d-\hat u) + p_2.n (\mu_\chi + \mu_u-\hat t)]
 + 2 \left[ ( v_V^f )^2 - ( a_V^f )^2 \right] \sqrt{\mu_u \mu_d}
(p_1.n+ p_2.n) \bigg) \non\\
&& + 2 \left[ ( G^L_{jiV} )^2 + ( G^R_{jiV} )^2 \right]  v_V^f a_V^f 
[-p_1.n (\mu_\chi+\mu_d-\hat u) + p_2.n (\mu_\chi + \mu_u-\hat t)]  \non\\
&&  +
4 G^L_{jiV} G^R_{jiV}
v_V^f a_V^f\sqrt{\mu_\chi} [-p_1.n (1+\mu_d-\hat t) + p_2.n 
(1+\mu_u-\hat u)] \Bigg\}
\eeq
\beq
\dx \Gamma^S_{\Phi} &=& 
\sum_k \frac{2}{(1+\mu_\chi + \mu_u + \mu_d - \mu_k - \hat u-\hat t)^2} 
\Bigg\{ 
\left[ ( G^L_{ijk} )^2 - ( G^R_{ijk} )^2 \right] 
[p_1.n + p_2.n]  
\non\\
&& \bigg(
\left[ ( v_k^f )^2 + ( a_k^f )^2 \right] [1+\mu_\chi-\hat u-\hat t] - 2
\left[ ( v_k^f )^2 - ( a_k^f )^2 \right] \sqrt{\mu_u\mu_d} \bigg) \Bigg\}
\eeq
\beq
\dx \Gamma^S_{H_1 H_2} &=& 
\frac{4 v_{H_1}^f v_{H_2}^f}
{(1+\mu_\chi + \mu_u + \mu_d - \mu_{H_1}- \hat u-\hat t)
(1+\mu_\chi + \mu_u + \mu_d - \mu_{H_2} - \hat u-\hat t)} \non\\
&& \left[ G^L_{ij1} G^L_{ij2} - G^R_{ij1} G^R_{ij2} \right] 
[p_1.n + p_2.n] (1+\mu_\chi-2\sqrt{\mu_u\mu_d}-\hat u-\hat t)
\eeq 

\beq
\dx \Gamma^S_{V\Phi} &=&  \sum_{k=1}^2 \frac{4}
{(1+\mu_\chi + \mu_u + \mu_d - \mu_{H_k}- \hat u-\hat t)
(1+\mu_\chi + \mu_u + \mu_d - \mu_{V} - \hat u-\hat t)} \non\\
&& \Bigg\{ \left[ G^L_{jiV} G^R_{ijk} - G^R_{jiV} G^L_{ijk} \right] 
\bigg( \left[v_k^f v_V^f + a_k^f a_V^f\right] 
\sqrt{\mu_u} [p_1.n (\hat t-1-\mu_d) 
+ p_2.n (-\hat u-1 \non\\
&& +\mu_u)]+ \left[v_k^f v_V^f - a_k^f a_V^f\right] \sqrt{\mu_d} 
[p_1.n (\hat t-\mu_d+1)
+ p_2.n (-\hat u+\mu_u+1)] \bigg) 
\non\\
&&  + 
2\left[ G^L_{jiV} G^L_{ijk} - G^R_{jiV} G^R_{ijk} \right]
\sqrt{\mu_\chi}
\bigg( -p_2.n\sqrt{\mu_u}\left[v_k^f v_V^f + a_k^f a_V^f\right] 
 \non\\
&& + p_1.n\sqrt{\mu_d}\left[v_k^f v_V^f - a_k^f a_V^f\right] \bigg) \Bigg\}
\eeq
\beq
\dx \Gamma^S_{\tilde u} = \sum_{k,l=1}^2 \frac{p_2.n}
{(- \mu_d - \mu_{\tilde u_k}+ \hat t)
(- \mu_d - \mu_{\tilde u_l} + \hat t)} \Bigg\{
2 a_{2S}^u \sqrt{\mu_u\mu_\chi} 
+ a_{4S}^u (\hat t-\mu_\chi-\mu_u) \Bigg\}
\eeq
\beq
\dx \Gamma^S_{\tilde d} = \sum_{k,l=1}^2  \frac{- p_1.n}
{(- \mu_d - \mu_{\tilde d_k}+ \hat u)
(- \mu_d - \mu_{\tilde d_l} + \hat u)} \Bigg\{
2 a_{2S}^d \sqrt{\mu_d\mu_\chi} 
+a_{4S}^d (\hat u-\mu_\chi-\mu_d) \Bigg\}
\eeq
where
\beq
a_{2S}^f &=& (a^f_{jk} b^f_{jl} + a^f_{jl} b^f_{jk})
(a^f_{ik} a^f_{il} - b^f_{ik} b^f_{il}) \non\\
a_{4S}^f &=& (a^f_{jk} a^f_{jl} + b^f_{jk} b^f_{jl})
(- a^f_{ik} a^f_{il} + b^f_{ik} b^f_{il})
\eeq
\beq
\dx \Gamma^S_{V\tilde d} &=& \sum_{l=1}^2 \frac{-2}
{(1+\mu_\chi + \mu_u + \mu_d - \mu_{V}- \hat u-\hat t)
(- \mu_u - \mu_{\tilde d_l} + \hat u)} \Bigg\{ 2 b_{1S}^{d}(jiV) 
p_1.n (\hat u -\mu_\chi -\mu_d ) \non\\
&& + b_{2S}^{d}(jiV) \sqrt{\mu_\chi} 
[p_1.n (\hat t-\mu_d-1) - p_2.n (\hat u-\mu_u-1)] 
- 2 b_{3S}^{d}(jiV) \sqrt{\mu_u \mu_d} (p_1.n 
\non\\
&& + p_2.n)
+ b_{5S}^{d}(jiV)\sqrt{\mu_d} 
[p_1.n (\hat t-\mu_d+1) - p_2.n (\hat u-\mu_u-1)] \non\\
&& -2 b_{6S}^{d}(jiV) p_2.n \sqrt{\mu_\chi \mu_u} -4 b_{8S}^{d}(jiV)
p_1.n \sqrt{\mu_\chi \mu_d} \Bigg\}
\eeq
\beq
\dx \Gamma^S_{V\tilde u} &=& \sum_{l=1}^2 \frac{2}
{(1+\mu_\chi + \mu_u + \mu_d - \mu_{V}- \hat u-\hat t)
(- \mu_d - \mu_{\tilde u_l} + \hat t)} \Bigg\{ b_{1S}^{u}(jiV) 
\sqrt{\mu_\chi}  \non\\
&& [p_1.n (\hat t-1-\mu_d) - p_2.n (\hat u -1-\mu_u)] + 2 b_{2S}^{u}(jiV)
p_2.n (-\hat t+\mu_\chi+\mu_u) \non\\
&& + 2 b_{4S}^{u}(jiV) \sqrt{\mu_u\mu_d} (p_1.n + p_2.n) + 
4 b_{5S}^{u}(jiV) \sqrt{\mu_u \mu_\chi} p_2.n + 2 b_{7S}^{u}(jiV) 
\sqrt{\mu_\chi \mu_d} p_1.n \non\\
&& + b_{8S}^{u}(jiV) \sqrt{\mu_u} [p_1.n (\hat t-\mu_d-1) - p_2.n (\hat u 
-\mu_u +1)] \Bigg\}
\eeq
with
\beq
b_{1S}^{f}(ijk) = a^f_{il} a^f_{jl} G^R_{ijk} (v_k^f + a_k^f) - b^f_{il} b^f_{jl} 
G^L_{ijk} (v_k^f - a_k^f) \non\\
b_{2S}^{f}(ijk) = a^f_{il} a^f_{jl} G^L_{ijk} (v_k^f + a_k^f) - b^f_{il} b^f_{jl} 
G^R_{ijk} (v_k^f - a_k^f) \non\\
b_{3S}^{f}(ijk) = a^f_{il} a^f_{jl} G^R_{ijk} (v_k^f - a_k^f) - b^f_{il} b^f_{jl} 
G^L_{ijk} (v_k^f + a_k^f) \non\\
b_{4S}^{f}(ijk) = a^f_{il} a^f_{jl} G^L_{ijk} (v_k^f - a_k^f) - b^f_{il} b^f_{jl} 
G^R_{ijk} (v_k^f + a_k^f) \non\\
b_{5S}^{f}(ijk) = a^f_{jl} b^f_{il} G^L_{ijk} (v_k^f + a_k^f) - a^f_{il} b^f_{jl} 
G^R_{ijk} (v_k^f - a_k^f) \non\\
b_{6S}^{f}(ijk) = a^f_{jl} b^f_{il} G^R_{ijk} (v_k^f - a_k^f) - a^f_{il} b^f_{jl} 
G^L_{ijk} (v_k^f + a_k^f) \non\\
b_{7S}^{f}(ijk) = a^f_{jl} b^f_{il} G^L_{ijk} (v_k^f - a_k^f) - a^f_{il} b^f_{jl} 
G^R_{ijk} (v_k^f + a_k^f) \non\\
b_{8S}^{f}(ijk) = a^f_{jl} b^f_{il} G^R_{ijk} (v_k^f + a_k^f) - a^f_{il} b^f_{jl} 
G^L_{ijk} (v_k^f - a_k^f) \non\\
\eeq
\beq
\dx \Gamma^S_{\tilde u\tilde d} &=& \sum_{k,l=1}^2 \frac{-1}
{(- \mu_u - \mu_{\tilde d_l} + \hat u)
(- \mu_d - \mu_{\tilde u_k} + \hat t)} \Bigg\{ 2[a^u_{ik} a^u_{jk} b^d_{il} 
b^d_{jl} - a^d_{il} a^d_{jl} b^u_{ik} b^u_{jk}] \sqrt{\mu_u\mu_d} 
 \non\\ 
&& [p_1.n + p_2.n] + [a^u_{ik} a^u_{jk} a^d_{il} b^d_{jl} -
b^u_{ik} b^u_{jk} b^d_{il} a^d_{jl}] \sqrt{\mu_d} [ p_1.n (\hat t-\mu_d +1)
- p_2.n (\hat u-\mu_u-1)] \non\\
&& -2 [a^u_{jk} a^d_{il} a^d_{jl} b^u_{ik} -
b^u_{jk} b^d_{il} b^d_{jl} a^u_{ik}] \sqrt{\mu_\chi\mu_d} p_1.n +
[a^u_{jk} b^u_{ik} b^d_{il} b^d_{jl} - a^u_{ik} a^d_{il} a^d_{jl} b^u_{jk}] 
\sqrt{\mu_u} [p_1.n  \non\\
&& (-\hat t+\mu_d+1)-p_2.n(-\hat u+\mu_u-1)] 
+ [a^u_{jk} a^d_{il} b^u_{ik} b^d_{jl} - a^u_{ik} a^d_{jl} b^u_{jk} b^d_{il}]
[p_1.n (-\hat t-\mu_d+1)  \non\\
&& +p_2.n (-\hat u-\mu_u+1) +
[a^u_{ik} a^u_{jk} a^d_{il} a^d_{jl} - b^u_{ik} b^u_{jk} b^d_{il} b^d_{jl}] 
\sqrt{\mu_\chi} [p_1.n(\hat t-\mu_d-1)-p_2.n(\hat u \non\\
&& -\mu_u-1)] + 2 [a^u_{ik} a^u_{jk} a^d_{jl} b^d_{il} 
- b^u_{ik} b^u_{jk} b^d_{jl} a^d_{il}] \sqrt{\mu_\chi\mu_u} p_2.n \Bigg\}
\eeq
\beq
\dx \Gamma^S_{\Phi_k \tilde d} &=& \sum_{k,l} \frac{1}
{(1+\mu_\chi + \mu_u + \mu_d - \mu_{\Phi_k}- \hat u-\hat t)
(- \mu_u - \mu_{\tilde d_l} + \hat u)} \Bigg\{ -2 b_{1S}^d(ijk)
\non\\
&& \sqrt{\mu_\chi\mu_u} p_2.n
+ b_{2S}^d(ijk)\sqrt{\mu_u} [p_1.n (\hat t-\mu_d-1) + p_2.n (-\hat u+\mu_u-1)] 
+ 2b_{3S}^d(ijk) \non\\
&& \sqrt{\mu_\chi\mu_d} p_1.n + b_{4S}^d(ijk) \sqrt{\mu_d} [p_1.n (\hat t-
\mu_d+1)+p_2.n (-\hat u+\mu_u+1)] + b_{5S}^d(ijk)\non\\
&& [p_1.n (2\hat u+\hat t-2\mu_\chi-\mu_d-1)+p_2.n (\hat u+\mu_u-1)] +
2 b_{7S}^d(ijk) \sqrt{\mu_u\mu_d} [p_1.n + p_2.n] \non\\
&& + b_{8S}^d(ijk) \sqrt{\mu_\chi} [p_1.n 
(-\hat t+\mu_d+1) + p_2.n (\hat u-\mu_u-1)] \Bigg\}
\eeq
\beq
\dx \Gamma^S_{\Phi_k \tilde u} &=& \sum_{k,l} \frac{1}
{(1+\mu_\chi + \mu_u + \mu_d - \mu_{\Phi_k}- \hat u-\hat t)
(- \mu_d - \mu_{\tilde u_l} + \hat t)} \Bigg\{ b_{1S}^u(ijk)
\non\\
&& \sqrt{\mu_u} [p_1.n (\hat t-\mu_d-1) + p_2.n (-\hat u+\mu_u-1)] 
-2 b_{2S}^u(ijk)\sqrt{\mu_\chi\mu_u} p_2.n + b_{3S}^u(ijk) \non\\
&& \sqrt{\mu_d} [p_1.n (\hat t-\mu_d+1) + p_2.n (-\hat u+\mu_u+1)] + 
2 b_{4S}^u(ijk) \sqrt{\mu_\chi\mu_d} p_1.n + b_{6S}^u(ijk)\non\\
&& [p_1.n (-\hat t-\mu_d+1)+p_2.n (-\hat u-2\hat t+2\mu_\chi+\mu_u+1)] 
+ b_7^u(ijk)\sqrt{\mu_\chi} [p_1.n (-\hat t+ \non\\
&& \mu_d +1)+ p_2.n (\hat u-\mu_u-1)]
-2 b_{8S}^d(ijk) \sqrt{\mu_u\mu_d} [p_1.n + p_2.n] \Bigg\}
\eeq

\subsection*{Phase space}

To obtain the integrated partial widths, one has to express $\hat{u}$ and
$\hat{t}$ as functions of $x_1$ and $x_2$ 
\beq
\hat{u}=1- x_1+ \mu_u \ \ ,  \ \
\hat{t}=1- x_2+ \mu_d 
\eeq
and integrate over the latter variables, with boundary conditions: 
\beq
2 \sqrt{\mu_u} \ \leq \ x_1 \ \leq 1+ [\mu_u -(\sqrt{\mu_d}+ \sqrt{\mu_\chi})
^2] 
\eeq
\beq
s_{\rm min} \ \leq x_2 \ \leq s_{\rm max} 
\eeq
\beq
s_{\rm min} &=& \frac{1}{2} \frac{(x_1-2)(x_1 -1 -\mu_d + \mu_\chi -
\mu_u ) - \sqrt{\Delta} }{1-x_1+ \mu_u } \non \\
s_{\rm max} &=& \frac{1}{2} \frac{(x_1-2)(x_1 -1 -\mu_d + \mu_\chi -
\mu_u ) +\sqrt{\Delta} }{1-x_1+\mu_u } 
\eeq
with
\beq
\Delta =   (\mu_u -x_1^2) \left[ \frac{1}{4} \mu_d \mu_\chi - (x_1 -1 +
\mu_d +\mu_\chi - \mu_u )^2 \right]
\eeq


\begin{thebibliography}{99}
\bibitem{sdecay} Y. Mambrini, {\tt SDECAY}: a program for the decays of SUSY 
particles, to be released.
\end{thebibliography}
\end{document}